\documentclass[12pt,twoside]{article}   
\usepackage[super,sort,comma]{natbib}

\usepackage{fancyhdr}		

\usepackage[section]{placeins}   %

\usepackage{graphicx}

\makeatletter \renewcommand\@biblabel[1]{$^{#1}$} \makeatother
\usepackage[mathlines]{lineno}

\usepackage{hyperref}
\hypersetup{ colorlinks,
	citecolor=blue,
	filecolor=blue,
	linkcolor=blue,
	urlcolor=blue
}

\usepackage{xcolor}

\definecolor{gray}{rgb}{0.6,0.6,0.6}
\definecolor{red}{rgb}{0.85,0,0}
\definecolor{green}{rgb}{0,0.85,0}
\definecolor{blue}{rgb}{0,0,0.85}
\definecolor{beige}{rgb}{0.92,0.87,0.78}
\definecolor{amaranth}{rgb}{0.9, 0.17, 0.31}
\usepackage[all]{hypcap}    

\usepackage{amsmath}
\usepackage{amssymb}
\usepackage[normalem]{ulem}
\usepackage{subfig}

\begin{document}
\begin{center}
{\Large {\bfseries The low-LET radiation contribution to the tumor dose in diffusing alpha-emitters radiation therapy}}  \\  
\vspace*{10mm}
{\large {\bfseries L. Epstein$^{1,2,3}$, G. Heger$^{1}$, A. Roy$^{1}$, I. Gannot$^{2}$, I. Kelson$^4$ and L. Arazi$^{1,*}$}}\\
\vspace{5mm}
{$^{1}$Unit of Nuclear Engineering, Faculty of Engineering Sciences, Ben-Gurion University of the Negev, Be'er-Sheva, Israel} 
\\ {$^{2}$Department of Biomedical Engineering, Faculty of Engineering, Tel Aviv University, Tel Aviv, Israel}\\
{$^{3}$Soreq Nuclear Research Center, Yavne, Israel} \\
{$^{4}$School of Physics and Astronomy, Faculty of Exact Sciences, Tel Aviv University, Tel Aviv, Israel }\\
\vspace{3mm}
Version typeset \today\\
\vspace{3mm}
$^*$Corresponding author: Lior Arazi, larazi@bgu.ac.il 
\end{center} 

\pagenumbering{roman}
\setcounter{page}{1}
\pagestyle{plain}

\begin{abstract}
\noindent {\bf Background:} Diffusing alpha-emitters Radiation Therapy (``Alpha DaRT'') is a new technique that enables the use of alpha particles for the treatment of solid tumors. Alpha DaRT employs interstitial sources carrying a few $\mu$Ci of $^{224}$Ra below their surface, designed to release a chain of short-lived atoms (progeny of $^{224}$Ra) which emit alpha particles, along with beta, Auger, and conversion electrons, x- and gamma rays. These atoms diffuse around the source and create---primarily through their alpha decays---a lethal high-dose region measuring a few millimeters in diameter.\\
{\bf Purpose:} While previous studies focused on the dose from the alpha emissions alone, this work addresses the electron and photon dose contributed by the diffusing atoms and by the atoms remaining on the source surface, for both a single Alpha DaRT source and multi-source lattices. This allows to evaluate the low-LET contribution to the tumor dose and tumor cell survival, and demonstrate the sparing of surrounding healthy tissue.\\
{\bf Methods:} The low-LET dose is calculated using the EGSnrc and FLUKA Monte Carlo codes. We compare the results of a simple line-source approximation with no diffusion to those of a full simulation, which implements a realistic source geometry and the spread of diffusing atoms. We consider two opposite scenarios: one with low diffusion and high $^{212}$Pb leakage, and the other with high diffusion and low leakage. The low-LET dose in source lattices is calculated by superposition of single-source contributions. Its effect on cell survival is estimated with the linear quadratic model in the limit of low dose rate.\\ 
{\bf Results:}  For sources carrying 3~$\mu$Ci/cm $^{224}$Ra arranged in a hexagonal lattice with 4~mm spacing, the minimal low-LET dose between sources is $\sim 18-30$~Gy for the two test cases and is dominated by the beta contribution. The low-LET dose drops below 5~Gy $\sim3$ mm away from the outermost source in the lattice with an effective maximal dose rate of $<0.04$~Gy/h. The accuracy of the line-source/no-diffusion approximation is $\sim15\%$ for the total low-LET dose over clinically relevant distances (2-4~mm). The low-LET dose reduces tumor cell survival by a factor of $\sim2-200$.\\ 
{\bf Conclusions:} The low-LET dose in Alpha DaRT can be modeled by conventional Monte Carlo techniques with appropriate leakage corrections to the source activity. For 3~$\mu$Ci/cm $^{224}$Ra sources, the contribution of the low-LET dose can reduce cell survival inside the tumor by up to two orders of magnitude. The low-LET dose to surrounding healthy tissue is negligible. Increasing source activities by a factor of 5 can bring the low-LET dose itself to therapeutic levels, in addition to the high-LET dose contributed by alpha particles, leading to a ``self-boosted'' Alpha DaRT configuration, and potentially allowing to increase the lattice spacing.\\
\end{abstract}




\setlength{\baselineskip}{0.7cm}      

\pagenumbering{arabic}
\setcounter{page}{1}

%
%
%

%
%

\section{Introduction}
The therapeutic potential of alpha particles in the treatment of cancer has long been recognized \cite{hall2018radiobiology,Raju1991,MIRD_Pamphlet_22}, leading to multiple clinical studies as Targeted Alpha Therapy (TAT) \cite{McDevitt2018,TAT_WG2018,Tafreshi2019}, and to the approved use of $^{223}$RaCl$_2$ treatments for bone metastases in castration-resistant prostate cancer\cite{Shirley2014}. Because of the short range of alpha particles, TAT is generally considered suitable for the treatment of single cells and micrometastatic disease, although recent results with PSMA-TAT show efficacy also against macroscopic metastases\cite{Sathekge2019}. In contrast, Diffusing alpha-emitters Radiation Therapy (``Alpha DaRT'') is a new modality which {\it a priori} focuses on the use of alpha particles against solid tumors. Previous publications have covered its basic principle and physics \cite{Arazi2007,Arazi2010,Arazi2020,Heger2023a, Heger2023b}, pre-clinical studies on mice-borne tumors as a stand-alone treatment\cite{Cooks2008,Cooks2009a,Cooks2012}, in combination with chemotherapy \cite{Cooks2009b,Horev-Drori2012,Reitkopf-Brodutch2015,Nishri2022}, anti-angiogenesis therapy\cite{Nishri2022} and immunotherapy \cite{Keisari2014,Confino2015,Confino2016,Domankevich2019,Domankevich2020,Keisari2020,Del_Mare2022}, as well as first clinical results \cite{Popovtzer2020,Bellia2019}. 

\begin{figure}
	\centering
	\includegraphics[width=0.5\textwidth]{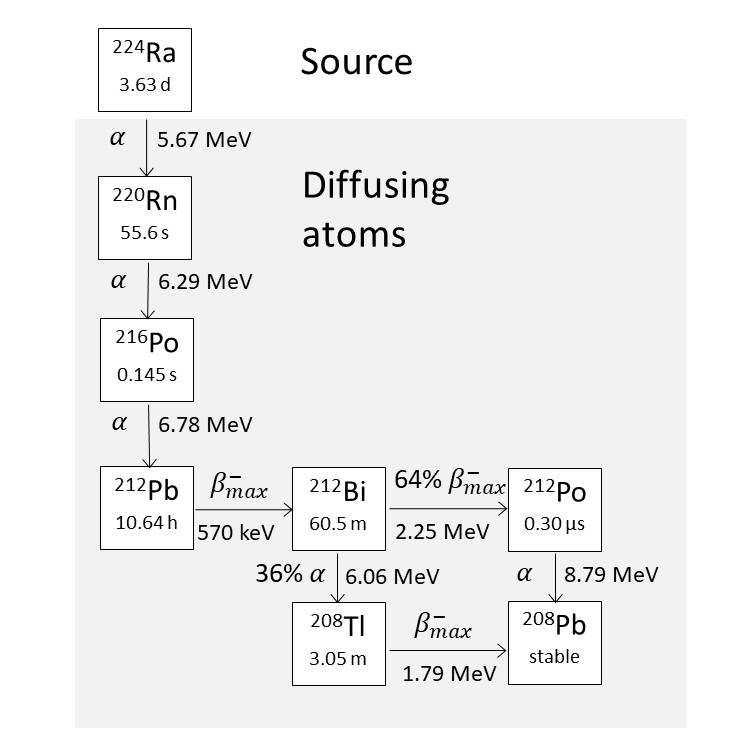}
	\caption{The $^{224}$Ra decay chain.}
	\label{fig:Fig1_RaDecayScheme}
\end{figure}

Alpha DaRT utilizes  radioactive sources embedded with a few $\mu$Ci of $^{224}$Ra, which are inserted into malignant tumors for a duration of at least two weeks. Once inside the tumor, the sources are designed to continuously release from their surface the short-lived daughters of $^{224}$Ra: $^{220}$Rn, $^{216}$Po, $^{212}$Pb, $^{212}$Bi, $^{212}$Po and $^{208}$Tl, ending with stable $^{208}$Pb (Figure \ref{fig:Fig1_RaDecayScheme}). The radioactive atoms disperse in the tumor mainly by diffusion, creating, through their radioactive emissions, a lethal high-dose region around each source with a diameter of typically 3-5 mm, where the alpha dose is 10 Gy or higher \cite{Arazi2007,Cooks2009a,Cooks2012,Horev-Drori2012,Reitkopf-Brodutch2015}. In the first-in-human trial \cite{Popovtzer2020}, focusing on locally advanced and recurrent squamous cell carcinoma of the skin, head and neck, tumors were treated with sources carrying 2 $\mu$Ci (74 kBq) $^{224}$Ra with a nominal spacing of 5 mm between sources. All 28 lesions treated showed positive response (30-100\% reduction in their longest diameter) and 22/28 (78.6\%) displayed complete response (i.e., disappeared on the macroscopic level). In all cases side effects were mild to moderate, with no indication of radiation-induced damage either locally or systemically.  

The size of the high-dose region around each source depends on the dispersal of the radioactive atoms. To describe this process inside the tumor, a simplified approach (``the Diffusion-Leakage (DL) model'') was developed \cite{Arazi2020}, aiming to serve as a zero-order approximation for treatment planning. This model also considers the leakage of $^{212}$Pb out of the tumor through the blood. It describes the migration of the atoms as being purely diffusive and was initially used to estimate the alpha radiation dose by approximating sources as lines, and summing the contribution of point-like segments along their length. A more recent work, implementing a finite-element scheme, extended this to two dimensions to describe the alpha dose for realistic source geometries \cite{Heger2023a}.

Past work focused on the alpha particle dose as the main contributor in Alpha DaRT tumor dosimetry. Here we present, for the first time, detailed calculations of the dose from the fast electrons and gamma/x rays (low-LET radiation) emitted by $^{224}$Ra and its daughters. The dose is calculated as a function of distance from a point source in water to establish dose point kernels (DPKs) of the low-LET emissions of $^{224}$Ra, $^{212}$Pb, $^{212}$Bi and $^{208}$Tl, using the EGSnrc\citep{Kawrakow2001,Kawrakow2021} and FLUKA\cite{Ahdida2022,Battistoni2015} Monte Carlo (MC) codes for electron and photon transport. These DPKs are used to calculate the total low-LET dose from an Alpha DaRT point source, including the contributions of both the atoms remaining on the source and those diffusing in its vicinity. We use the point source expressions to understand the main factors affecting the low-LET dose and show that the effect of diffusion can be neglected to first order. The analysis is then extended to a realistic Alpha DaRT source geometry using a full MC calculation by FLUKA, which accounts for the spatial spread of the diffusing atoms. We show that a line-source/no-diffusion model leads to a reasonable first-order approximation to the full MC calculation, provided that $^{212}$Pb leakage is accounted for as an effective reduction in the release rate of $^{224}$Ra daughters from the source. We use the full MC calculation as a basis for source lattice calculations and investigate the low-LET dose inside the lattice and in the tumor periphery. Lastly, we consider the effect of the low-LET dose on cell survival and discuss how boosting the low-LET contribution can improve tumor control without increasing the systemic dose, while potentially allowing for increased source spacing.

\section{Single-source calculations}
\subsection{Beta and gamma/x-ray emission data}
The $^{224}$Ra decay scheme is displayed in Figure \ref{fig:Fig1_RaDecayScheme}, with the maximal endpoint energies given for all beta emitters. The doses in the following sections were calculated for the beta electrons emitted by $^{212}$Pb, $^{212}$Bi and $^{208}$Tl and the gamma/x-ray emissions of $^{224}$Ra, $^{212}$Pb, $^{212}$Bi and $^{208}$Tl. The ratio between the total energy deposited by conversion and Auger electrons and that deposited by beta emissions is 68.2\% for $^{212}$Pb, 2.0\% for $^{212}$Bi and 6.5\% for $^{208}$Tl, as listed in the Nudat3 database\citep{Nudat3} (with average values of energy per decay of 69~keV for $^{212}$Pb, 4~keV for $^{212}$Bi, and 36~keV for $^{208}$Tl). For $^{224}$Ra the electron contribution is completely negligible. Since our focus is on the total low-LET dose, and---as shown below---the contribution of $^{212}$Pb is smaller by several orders of magnitude compared to $^{212}$Bi and $^{208}$Tl, we did not include Auger and conversion electrons in the present calculations. The beta spectra, gamma- and x-ray data were taken from the ENDF/B-VIII.0 library as published in JANIS \citep{Soppera2014, osti_1425114}. Only gamma and x-ray emissions with intensities $>1\%$ were considered:
\begin{itemize}
	\item $^{224}$Ra: 241.0~keV (4.1\%).
	\item $^{212}$Pb: 10.8~keV (6.7\%), 13.0~keV (5.7\%), 15.4~keV (1.1\%), 75.1~keV (10.0\%), 77.4~keV (16.8\%), 87.1~keV (1.9\%), 87.6~keV (3.6\%), 90.0~keV (1.4\%), 238.6~keV (43.6\%) and 300.1~keV (3.3\%).
	\item $^{212}$Bi: 10.3~keV (3.7\%), 12.2~keV (2.9\%) 39.9~keV (1.1\%), 727.3~keV (6.7\%), 785.4~keV (1.1\%) and 1620.5~keV (1.5\%).
	\item $^{208}$Tl: 10.5~keV (1.4\%), 12.6~keV (1.2\%), 73.0~keV (2.0\%), 75.2~keV (3.4\%), 277.4~keV (6.6\%), 510.8~keV (22.6\%), 583.2~keV (85.0\%), 763.1~keV (1.8\%), 860.6~keV (12.5\%) and 2614.5~keV (99.8\%). 
\end{itemize}

\begin{figure}
	\centering
	\includegraphics[width=1\textwidth]{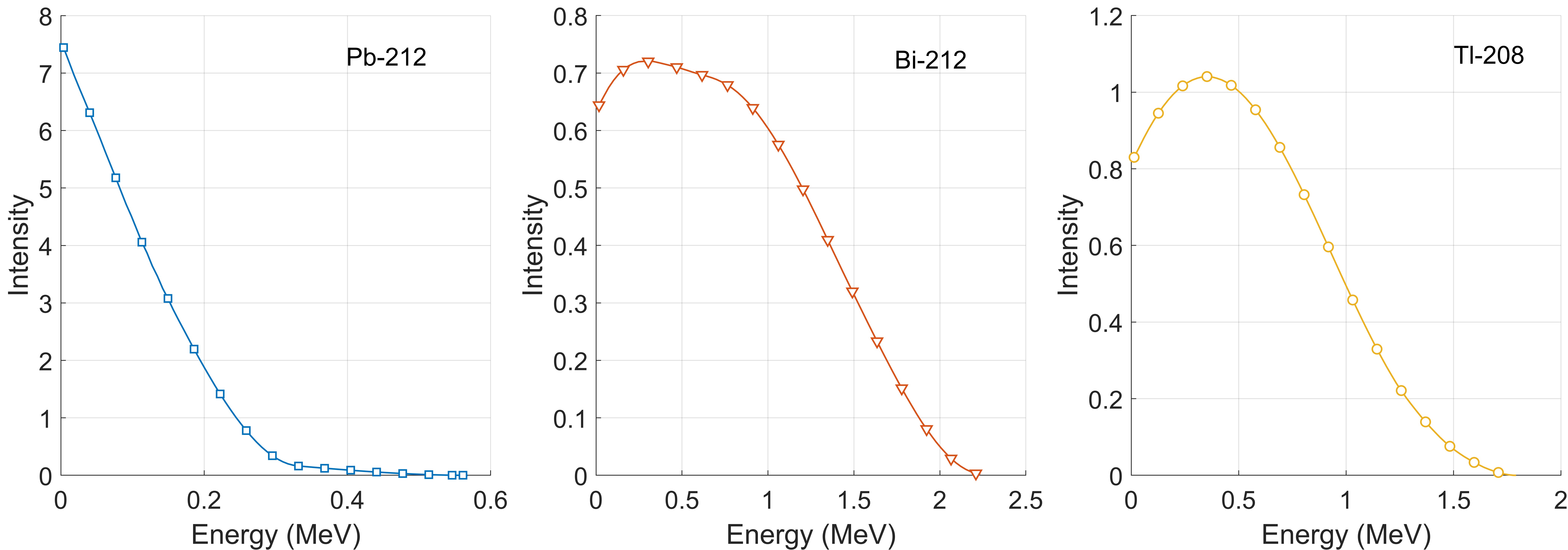}
	\caption{The beta spectra of $^{212}$Pb, $^{212}$Bi and $^{208}$Tl. The curves are normalized to an integrated area of 1. }
	\label{fig:Fig2_BetaSpectra}
\end{figure}

\noindent{The normalized beta spectra of $^{212}$Pb, $^{212}$Bi, and $^{208}$Tl are presented in Figure \ref{fig:Fig2_BetaSpectra}.}

\subsection{Dose point kernels}
Dose point kernels (DPKs) are routinely used in medical applications of radioactive sources. Conventionally, DPKs are found using MC simulations by calculating the dose as a function of distance from a point source in water\citep{Janicki2004,Jayarathna2019}. Here, we calculate the beta and gamma/x-ray DPKs using both the EGSnrc\citep{Kawrakow2001,Kawrakow2021} and  FLUKA\cite{Ahdida2022,Battistoni2015} MC codes. The DPKs are subsequently used as a basis for calculating the total low-LET dose from a point Alpha DaRT source.

The calculation was performed separately for each isotope and type of emitted particle, with a total of 7 simulations---four gamma/x-ray emitting isotopes: $^{224}$Ra, $^{212}$Pb, $^{212}$Bi, $^{208}$Tl, and three beta emitting isotopes: $^{212}$Pb, $^{212}$Bi, $^{208}$Tl. The contribution of each decay product to the dose was considered according to the branching ratio and the decay probability. 

The FLUKA calculations of the beta dose from $^{212}$Pb, $^{212}$Bi and $^{208}$Tl were done with a minimum of 5$\cdot$10$^8$ histories per isotope. The number of histories was determined according to the calculation statistics. A statistical error lower than 10\% was considered acceptable and most results are reported with a statistical error of less than 1\%.  The minimum energy threshold for the electron transport was set to 10 keV and to 1 keV for photons. The threshold was selected according to the particle's range or mean free path in water, such that it does not exceed the thickness of the dose calculation interval. The dose from a point source in water (density of 1.0 g$\cdot$cm$^{-3}$) was calculated as a function of radial distance $r$ up to 15~mm with a 100~$\mu$m grid in cylindrical coordinates $\rho-z$ (i.e., $\Delta\rho=\Delta z=0.1$~mm). FLUKA calculates the dose per beta decay at a radial distance $r$ from the source by dividing the total energy deposited in the associated ring at $\rho,z$ with widths $\Delta\rho,\Delta z$ ($r=\sqrt{\rho^2+z^2}$) by the ring mass and the total number of histories. 

EGSnrc calculations of the dose per beta decay were performed in spherical coordinates with 5$\cdot$10$^7$ histories for each isotope. A statistical error lower than 10\% was considered acceptable and most results are reported with a statistical error of less than 1\%. The minimum energy threshold was set as 
1 keV for photons and 10 keV for electrons but the cross section data included in the EGS tables has an internal energy threshold of 10 keV for both particle types. The dose was calculated as a function of distance from the source up to 15~mm with a 200~$\mu$m radial grid.

\begin{figure}
	\centering
	\includegraphics[scale=0.2,width=0.49\textwidth]{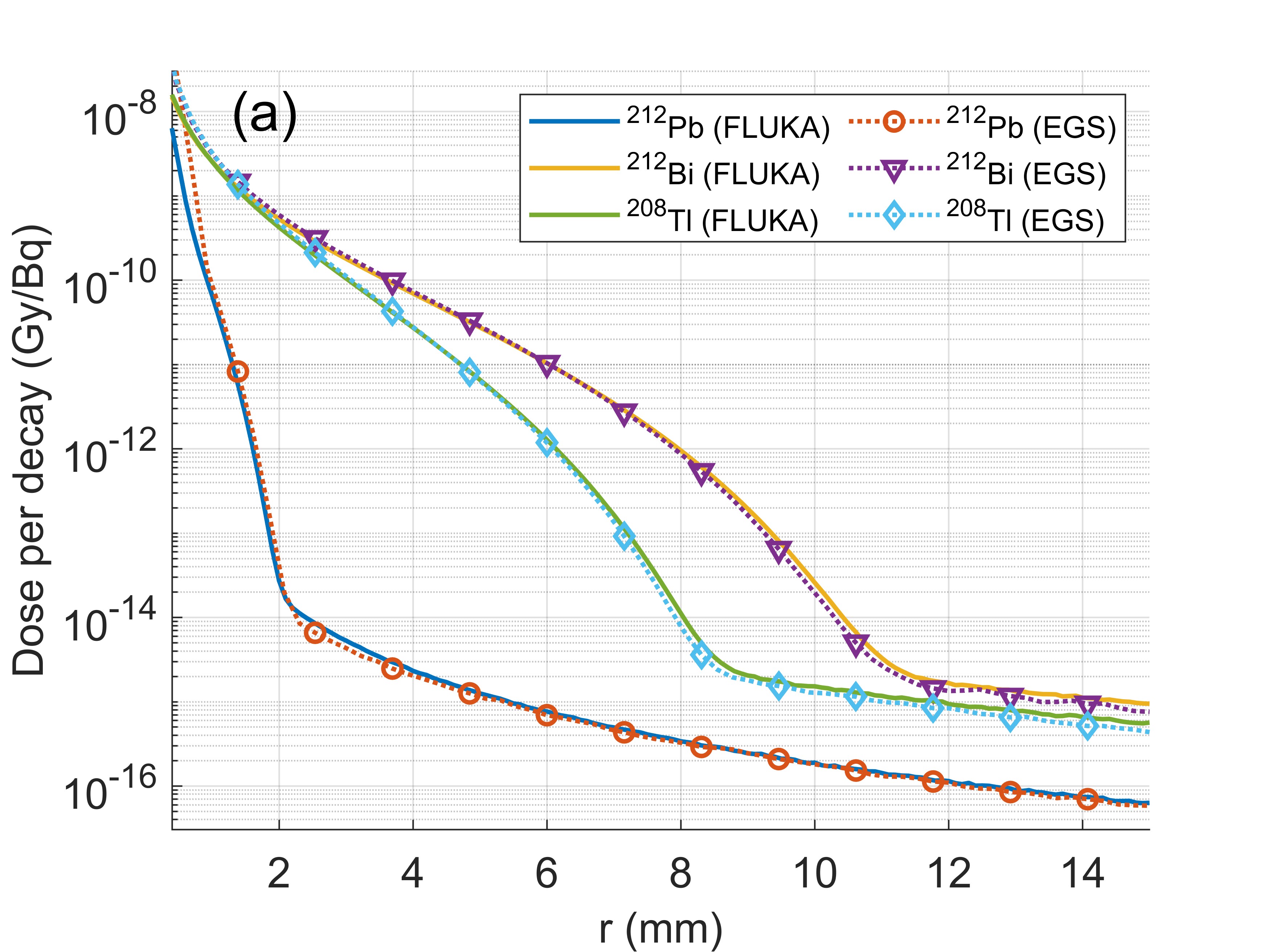}
	\includegraphics[scale=0.2,width=0.49\textwidth]{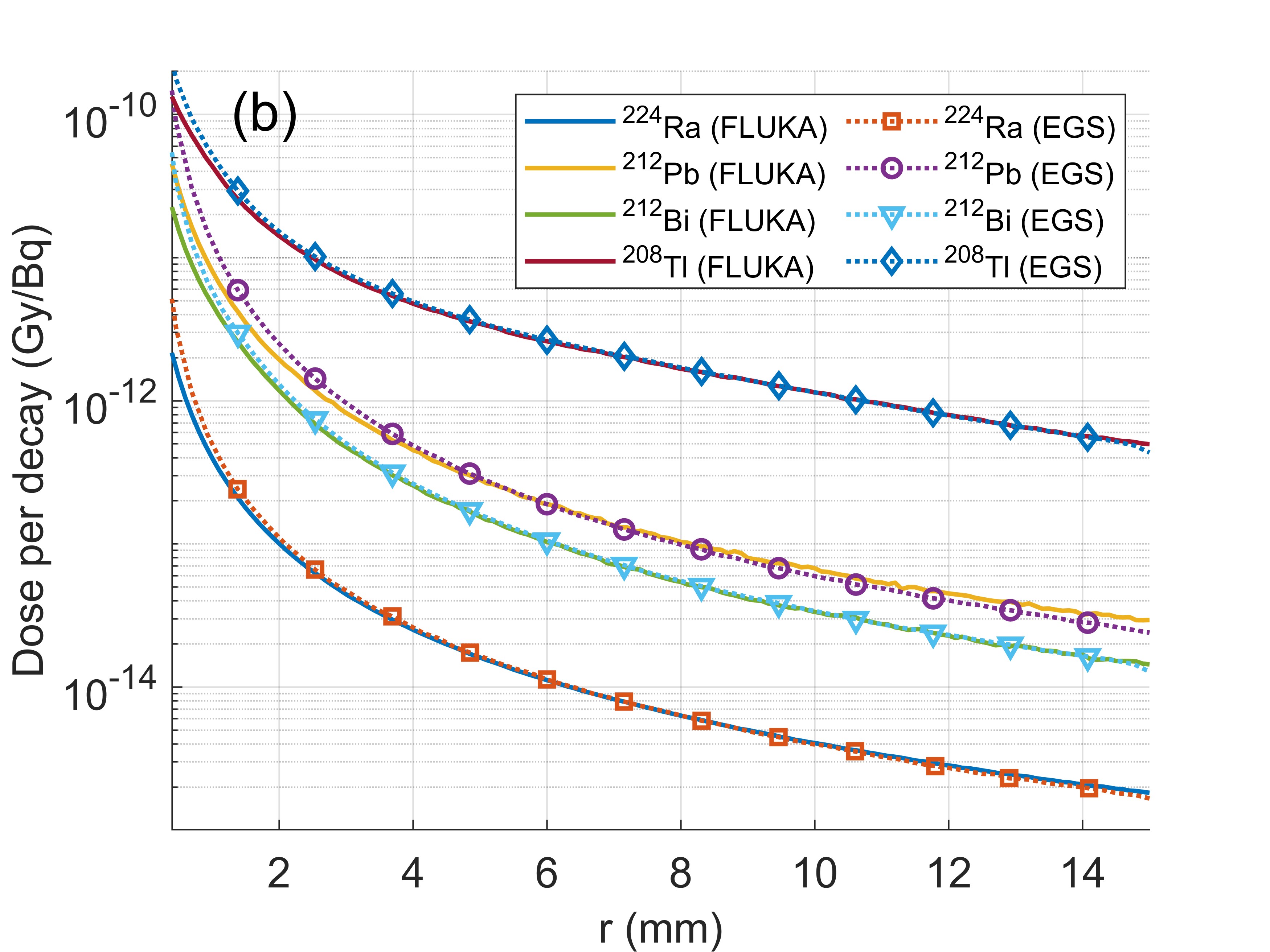}
	\caption{Dose point kernels (DPKs) calculated using FLUKA and EGSnrc. (a) Beta DPKs for $^{212}$Pb, $^{212}$Bi and $^{208}$Tl; (b) Gamma/x-ray DPKs for $^{224}$Ra, $^{212}$Pb, $^{212}$Bi and $^{208}$Tl. The beta DPKs are defined here as the dose per beta decay, while the gamma/x-ray DPKs are the dose per {\it radioactive} decay. }
	\label{fig:DPKs}
\end{figure}

Figure~\ref{fig:DPKs}a presents the beta DPKs as calculated by FLUKA and EGSnrc. The dose is dominated by the decays of $^{212}$Bi and $^{212}$Tl, with similar contributions at therapeutically-relevant distances ($\sim2-3$~mm) from the source. For each isotope, the flat tail at large distances (above the ``knee''), arises from energy deposition by bremsstrahlung photons. The relative difference between the two codes for $^{212}$Bi and $^{208}$Tl is $\sim6-11\%$ at a radial distance of $2-3$~mm; for $^{212}$Pb, whose contribution is smaller by $\sim4$ orders of magnitude, the relative difference over the same range is $\sim15-20\%$.

The gamma/x-ray DPKs, shown in Figure~\ref{fig:DPKs}b, were calculated similarly to the beta DPKs. Unlike the beta DPKs, which are defined here as the dose per beta decay, the gamma/x-ray DPKs are defined as the dose per {\it radioactive} decay, with the individual photon emissions sampled according to their intensities. The gamma/x-ray dose is dominated by the 2.615~MeV gamma of $^{208}$Tl. At a distance of $\sim2-3$~mm from the source it is $>30$-fold smaller than the beta dose, and becomes the dominant low-LET term only for $r\gtrsim8$~mm, where---as we show below---the absolute dose is negligibly small. The relative difference between the codes is $\sim5-8\%$ for $^{208}$Tl and $\sim17-30\%$ for $^{212}$Pb at $r=2-3$~mm.

\subsection{Low-LET dose from an Alpha DaRT point source}
Both the beta and gamma/x-ray doses from an Alpha DaRT source are the sum of contributions from emissions by atoms located on the source itself ({\it src}) and by atoms which have been released from the source and diffuse in its vicinity ({\it dif}):
\begin{equation}
	Dose_{\beta}(\mathbf{r},t)=Dose_{\beta}^{src}(\mathbf{r},t)+Dose_{\beta}^{dif}(\mathbf{r},t) \\
	\label{eq:totBeta}
\end{equation}
\begin{equation}
	Dose_{\gamma}(\mathbf{r},t)=Dose_{\gamma}^{src}(\mathbf{r},t)+Dose_{\gamma}^{dif}(\mathbf{r},t)
	\label{eq:totGamma}	
\end{equation}

As done previously for the alpha dose\cite{Arazi2020}, we begin with an Alpha DaRT point source as the basic building block for an arbitrary configuration of line sources, which---as we show below---can provide a first-order approximation for realistic Alpha DaRT sources when calculating the low-LET. For the point source, the 3D position vector $\mathbf{r}$ appearing in Equations (\ref{eq:totBeta}) and (\ref{eq:totGamma}) is simply replaced by the radial distance from the point source $r$.

\subsubsection{Contribution from the source to the beta and gamma/x-ray dose}

The beta dose at a radial distance $r$ from the point source at time $t$ (where $t=0$ is the time of source insertion into the tumor) is calculated by summing contributions from the decays of $^{212}$Pb, $^{212}$Bi and $^{208}$Tl on the source:
\begin{equation}
	\label{eq:betaDose_point_source_t}
	Dose_{\beta}^{src}(r,t)=\int_{0}^{t} \left(\Gamma_{Pb}^{src}(t')f^{\beta}_{Pb}(r)+0.64\Gamma_{Bi}^{src}(t')f^{\beta}_{Bi}(r)+\Gamma_{Tl}^{src}(t')f^{\beta}_{Tl}(r)\right) dt' 
\end{equation}
\noindent where $\Gamma_{Pb}^{src}(t')$, $\Gamma_{Bi}^{src}(t')$ and $\Gamma_{Tl}^{src}(t')$ are the $^{212}$Pb, $^{212}$Bi and $^{208}$Tl activities on the source at time $t'$, and  $f^{\beta}_{Pb}(r)$, $f^{\beta}_{Bi}(r)$ and $f^{\beta}_{Tl}(r)$ are their respective beta DPKs. This expression accounts for the 64\% branching ratio of the $^{212}$Bi beta decay. Defining the asymptotic dose as the dose from source insertion to infinity (in practice several weeks):
\begin{multline}
	\label{eq:betaDose_point_source_asy}
	Dose_{\beta}^{asy,src}(r)=\int_{0}^{\infty} \left(\Gamma_{Pb}^{src}(t')f^{\beta}_{Pb}(r)+0.64\Gamma_{Bi}^{src}(t')f^{\beta}_{Bi}(r)+\Gamma_{Tl}^{src}(t')f^{\beta}_{Tl}(r)\right)dt' \\
	= N_{decays}^{src}(Pb)f^{\beta}_{Pb}(r)+0.64N_{decays}^{src}(Bi)f^{\beta}_{Bi}(r)+N_{decays}^{src}(Tl)f^{\beta}_{Tl}(r)
\end{multline}
\noindent where $N_{decays}^{src}(Pb)$, $N_{decays}^{src}(Bi)$ and $N_{decays}^{src}(Tl)$ are the total number of decays of $^{212}$Pb, $^{212}$Bi and $^{208}$Tl, respectively, on the source. The $^{212}$Pb activity on the source at time $t$ is given by\cite{Arazi2020} :
\begin{equation} \label{eq:Pb_activity_src}
	\Gamma_{Pb}^{src}(t)=\Gamma_{Pb}^{src}(0)e^{-\lambda_{Pb}t}+\frac{\lambda_{Pb}}{\lambda_{Pb}-\lambda_{Ra}}\left(1-P^{eff}_{des}(Pb)\right)\Gamma_{Ra}^{src}(0)(e^{-\lambda_{Ra}t}-e^{-\lambda_{Pb}t})    
\end{equation}
where $\lambda_{Pb}$ and $\lambda_{Ra}$ are the respective decay rate constants of $^{212}$Pb and $^{224}$Ra, $\Gamma_{Ra}^{src}(0)$ and $\Gamma_{Pb}^{src}(0)$ are the initial $^{224}$Ra and $^{212}$Pb activities on the source, and $P^{eff}_{des}(Pb)$ is the {\it effective desorption probability} of $^{212}$Pb, i.e., the probability that a decay of a $^{224}$Ra atom on the source results in the release of one atom of $^{212}$Pb into the tumor, including prior release of either $^{220}$Rn or $^{216}$Po from the source (typically $P^{eff}_{des}(Pb)=0.55$) \cite{Arazi2020}. The initial $^{224}$Ra activity can be adjusted according to clinical needs and is usually 3~$\mu$Ci. 

\noindent The total number of $^{212}$Pb decays on the source is therefore:
\begin{equation}
	N_{decays}^{src}(Pb)=\int_0^{\infty}\Gamma_{Pb}^{src}(t)dt=\Gamma_{Pb}^{src}(0)\tau_{Pb}+\left(1-P^{eff}_{des}(Pb)\right)\Gamma_{Ra}^{src}(0)\tau_{Ra}
	\label{eq:N_Pb}
\end{equation}
where $\tau_{Pb}=1/\lambda_{Pb}$ and $\tau_{Ra}=1/\lambda_{Ra}$ are the mean lifetimes of $^{212}$Pb and $^{224}$Ra, respectively. Each decay of $^{212}$Pb gives rise to one atom of $^{212}$Bi, which, due to the low recoil energy in beta decay (few eV) can be assumed to remain on the source. 
Lastly, $^{208}$Tl is created through the alpha decay of $^{212}$Bi, which has a branching ratio of 36\%. Although it does recoil from the source into the tumor, it has a short half-live (3.05 min) and likely forms chemical bonds with surrounding molecules, with a small effective diffusion coefficient compared to a free ion. We therefore assume that $^{208}$Tl generated by alpha decays of $^{212}$Bi on the source essentially decays on the source itself, such that $N_{decays}^{src}(Tl)=0.36N_{decays}^{src}(Bi)$. With this, the asymptotic beta dose coming directly from the source is:
\begin{multline}\label{eq:BetaDoseSrc_point_source}
		Dose^{asy,src}_{\beta}(r) =\\
		\left[\Gamma_{Pb}^{src}(0)\tau_{Pb}+\left(1-P^{eff}_{des}(Pb)\right)\Gamma_{Ra}^{src}(0)\tau_{Ra}\right]\cdot \left(f^{\beta}_{Pb}(r)+0.64f^{\beta}_{Bi}(r)+0.36f^{\beta}_{Tl}(r)\right)   
\end{multline}

When received at the hospital, the source is contained in an applicator needle filled with glycerin, which serves two purposes: (1) prevention of unwanted $^{220}$Rn release from the applicator prior to the treatment, and (2) retention of the source inside the needle by viscous friction. Typically, the time from source production to the treatment is long compared to the half-life of $^{212}$Pb, such that $^{212}$Pb and $^{224}$Ra are in secular equilibrium inside the applicator, and $\Gamma_{Pb}^{tot}(0)\approx\frac{\lambda_{Pb}}{\lambda_{Pb}-\lambda_{Ra}}\Gamma_{Ra}^{src}(0)=1.14\,\Gamma_{Ra}^{src}(0)$. Here, $\Gamma_{Pb}^{tot}$ is the total $^{212}$Pb activity, including the activity on the source itself and inside the glycerin. When the source is inserted into the tumor some of the glycerin filling the needle accompanies it. For simplicity, we associate the $^{212}$Pb activity contained in the glycerin which enters the tumor with the initial $^{212}$Pb on the source, $\Gamma_{Pb}^{src}(0)$. Since the exact amount of glycerin activity is unknown, we consider two extreme cases: (1) the entire activity of $^{212}$Pb contained in the glycerin enters the tumor, such that $\Gamma_{Pb}^{src}(0)=1.14\,\Gamma_{Ra}^{src}(0)$; (2) No glycerin enters the tumor, such that $\Gamma_{Pb}^{src}(0)=1.14\,(1-P_{des}^{eff}(Pb))\Gamma_{Ra}^{src}(0)\approx0.51\,\Gamma_{Ra}^{src}(0)$. From eq. (\ref{eq:BetaDoseSrc_point_source}), the source beta dose in the second case (no glycerin contribution) is 0.90 times its value in the first case (full glycerin contribution). Assuming a nominal case halfway between these two extremes, the associated uncertainty in the source beta dose is $\sim5\%$. Assuming further that the source contributes $\sim60-80\%$ of the total beta dose to the tumor, the associated uncertainty in the total low-LET dose is $\sim3\%$.

\noindent{Similarly to the beta dose, the asymptotic gamma/x-ray dose from the source is:}
\begin{multline}\label{eq:GammaDoseSrc_point_source}
	Dose^{asy,src}_{\gamma}(r)=\Gamma_{Ra}^{src}(0)\tau_{Ra}f^{\gamma}_{Ra}(r)+\\
	\left[\Gamma_{Pb}^{src}(0)\tau_{Pb}+\left(1-P^{eff}_{des}(Pb)\right)\Gamma_{Ra}^{src}(0)\tau_{Ra}\right]\cdot \Bigl(f^{\gamma}_{Pb}(r)+f^{\gamma}_{Bi}(r)+0.36f^{\gamma}_{Tl}(r)\Bigr)    
\end{multline}
where $f^{\gamma}_{Ra}(r)$, $f^{\gamma}_{Pb}(r)$, $f^{\gamma}_{Bi}(r)$ and $f^{\gamma}_{Tl}(r)$ are the gamma/x-ray DPKs of $^{224}$Ra, $^{212}$Pb, $^{212}$Bi and $^{208}$Tl, respectively. Here, as well, reasonable variations in $\Gamma_{Pb}^{src}(0)/\Gamma_{Ra}^{src}(0)$ lead to variations of a few \% in the photon dose arising from the source itself.

\subsubsection{Contribution from the diffusing atoms to the beta and gamma/x-ray dose} \label{sec:point source diffusing atoms}
The contribution of the diffusing atoms to the beta dose at a point $\mathbf{r}$ is calculated by integrating over all space, summing the dose contributed from infinitesimal volume elements. Designating by $\mathbf{r'}$ the running integration point, we have:
\begin{multline}
			Dose^{asy,dif}_{\beta}(\mathbf{r})=\int_{t=0}^{\infty}\int_{\Omega} \Bigl\{\lambda_{Pb}n_{Pb}(\mathbf{r'},t)f^{\beta}_{Pb}(|\mathbf{r-r'}|) \\
			+0.64\lambda_{Bi}n_{Bi}(\mathbf{r'},t)f^{\beta}_{Bi}(|\mathbf{r-r'}|)+\lambda_{Tl}n_{Tl}(\mathbf{r'},t)f^{\beta}_{Tl}(|\mathbf{r-r'}|)\Bigr\} d^3\mathbf{r'} dt
\end{multline}
where the integration is over all space (the entire tumor, formally designated by ``$\Omega$'') and $|\mathbf{r-r'}|=R$ is the distance between the fixed point $\mathbf{r}$ at which the dose is sought and the integration point $\mathbf{r'}$. While the local specific activities of $^{212}$Pb and $^{212}$Bi are calculated individually, we assume that $^{208}$Tl is in local secular equilibrium with $^{212}$Bi: $\lambda_{Tl}n_{Tl}({\mathbf{r'},t})=0.36\lambda_{Bi}n_{Bi}({\mathbf{r'},t})$. Thus, for the asymptotic dose we have:
\begin{multline}
	Dose^{asy,dif}_{\beta}(\mathbf{r})=\int_{t=0}^{\infty}\int_{\Omega} \Bigl\{\lambda_{Pb}n_{Pb}(\mathbf{r'},t)f^{\beta}_{Pb}(|\mathbf{r-r'}|) \\
	+\lambda_{Bi}n_{Bi}(\mathbf{r'},t)\Bigl[0.64f^{\beta}_{Bi}(|\mathbf{r-r'}|)+0.36f^{\beta}_{Tl}(|\mathbf{r-r'}|)\Bigr]\Bigr\} d^3\mathbf{r'} dt
\end{multline}

\begin{figure}
	\centering
	\includegraphics[width=0.5\textwidth]{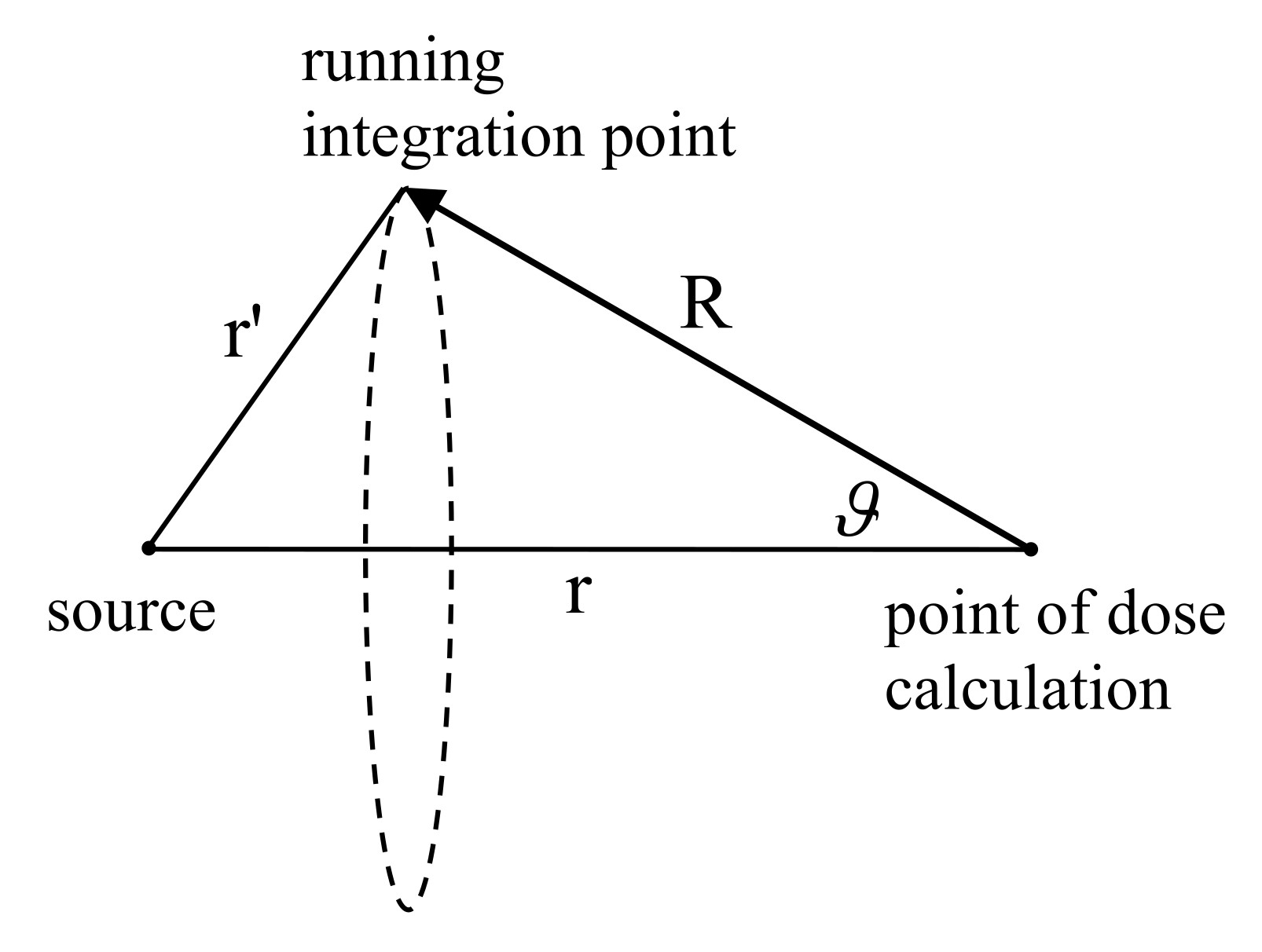}
	\caption{Integration scheme for calculating the beta dose from the diffusing atoms for the point source geometry.}
	\label{fig:Integration_scheme}
\end{figure}

For the point source the integration is carried out as illustrated in Figure \ref{fig:Integration_scheme}. We fix the origin for the integration at the point $\mathbf{r}$ (whose distance from the source is $r$). We scan the entire space at increasing distances from the point $\mathbf{r}$ with the vector $\mathbf{R}$ (with $R = |\mathbf{R}|$ running from 0 to $\infty$). Because of rotational symmetry about the line passing through the source and the point $\mathbf{r}$ the integration is carried out over infinitesimal rings (coaxial with the line of symmetry), defined by $R$ and $\vartheta$, where $\vartheta$ is the angle between the vector $\mathbf{R}$ and the symmetry axis. The distance from any point on the ring $(R,\vartheta)$ to the source is $r'=\sqrt{R^2+r^2-2Rr\textrm{cos}\vartheta}$. The volume of the infinitesimal ring $(R,\vartheta)$ is $dV_{ring} = 2\pi R^2\textrm{sin}\vartheta\,dR\,d\vartheta$. Thus, the
integral becomes:
\begin{multline} \label{eq:Dose_beta_point_source_triple_integral}
	Dose^{asy,dif}_{\beta}(r)=\int_{t=0}^{\infty}\int_{R=0}^{\infty}\int_{\vartheta=0}^{\pi} \Bigl\{\lambda_{Pb}n_{Pb}(r',t)f^{\beta}_{Pb}(R) \\
	+\lambda_{Bi}n_{Bi}(r',t)\Bigl[0.64f^{\beta}_{Bi}(R)+0.36f^{\beta}_{Tl}(R)\Bigr]\Bigr\} 2\pi R^2\textrm{sin}\vartheta\,dR\,d\vartheta\,dt
\end{multline}
The asymptotic gamma/x-ray dose contributed by the diffusing atoms is similarly:
\begin{multline} \label{eq:Dose_gamma_point_source_triple_integral}
	Dose^{asy,dif}_{\gamma}(r)=\int_{t=0}^{\infty}\int_{R=0}^{\infty}\int_{\vartheta=0}^{\pi} \Bigl\{\lambda_{Pb}n_{Pb}(r',t)f^{\gamma}_{Pb}(R) \\
	+\lambda_{Bi}n_{Bi}(r',t)\Bigl[f^{\gamma}_{Bi}(R)+0.36f^{\gamma}_{Tl}(R)\Bigr]\Bigr\} 2\pi R^2\textrm{sin}\vartheta\,dR\,d\vartheta\,dt
\end{multline}

The local $^{212}$Pb and $^{212}$Bi number densities appearing in Equations (\ref{eq:Dose_beta_point_source_triple_integral}) and (\ref{eq:Dose_gamma_point_source_triple_integral}), $n_{Pb}(r',t)$ and $n_{Bi}(r',t)$, can be either calculated numerically by solving the time dependent equations of the diffusion-leakage model, or approximated by assuming a uniform (``0D'') time dependence throughout the tumor, which does not consider the radial dependence of the delayed buildup of $^{212}$Pb away from the source\cite{Arazi2020}. Another approximation which---as we show below---provides a good estimate of the dose up to a few mm for the source is the following. We assume that  $^{212}$Pb atoms that leave the source (either directly or following the prior release of $^{220}$Rn or $^{216}$Po) may leak out of the tumor with a probability $P_{leak}(Pb)$ before their decay\cite{Arazi2020}, but---if they remain inside the tumor---diffuse to negligible distances from the source; we further assume that the resulting $^{212}$Bi and $^{208}$Tl atoms decay at the same location as $^{212}$Pb. In this ``no diffusion'' approximation the asymptotic beta and gamma/x-ray doses of the diffusing atoms are replaced by:
\begin{multline} \label{eq:Dose_beta_no_diff}
	Dose^{asy,no-dif}_{\beta}(r)=P_{des}^{eff}\Gamma_{Ra}^{src}(0)\tau_{Ra}\left(1-P_{leak}(Pb)\right)\cdot\left(f^{\beta}_{Pb}(r)+0.64f^{\beta}_{Bi}(r)+0.36f^{\beta}_{Tl}(r)\right)
\end{multline}
and
\begin{multline} \label{eq:Dose_gamma_no_diff}
	Dose^{asy,no-dif}_{\gamma}(r)=P_{des}^{eff}\Gamma_{Ra}^{src}(0)\tau_{Ra}\left(1-P_{leak}(Pb)\right)\cdot\Bigl(f^{\gamma}_{Pb}(r)+f^{\gamma}_{Bi}(r)+0.36f^{\gamma}_{Tl}(r)\Bigr)
\end{multline}
while the dose contribution of the source remains the same as in Equations (\ref{eq:BetaDoseSrc_point_source}) and (\ref{eq:GammaDoseSrc_point_source}). Note that the product $P_{des}^{eff}\Gamma_{Ra}^{src}(0)\tau_{Ra}\left(1-P_{leak}(Pb)\right)$ is the total number of $^{212}$Pb decays outside of the source but inside the tumor.

\begin{figure}
	\centering
	\includegraphics[scale=0.2,width=0.49\textwidth]{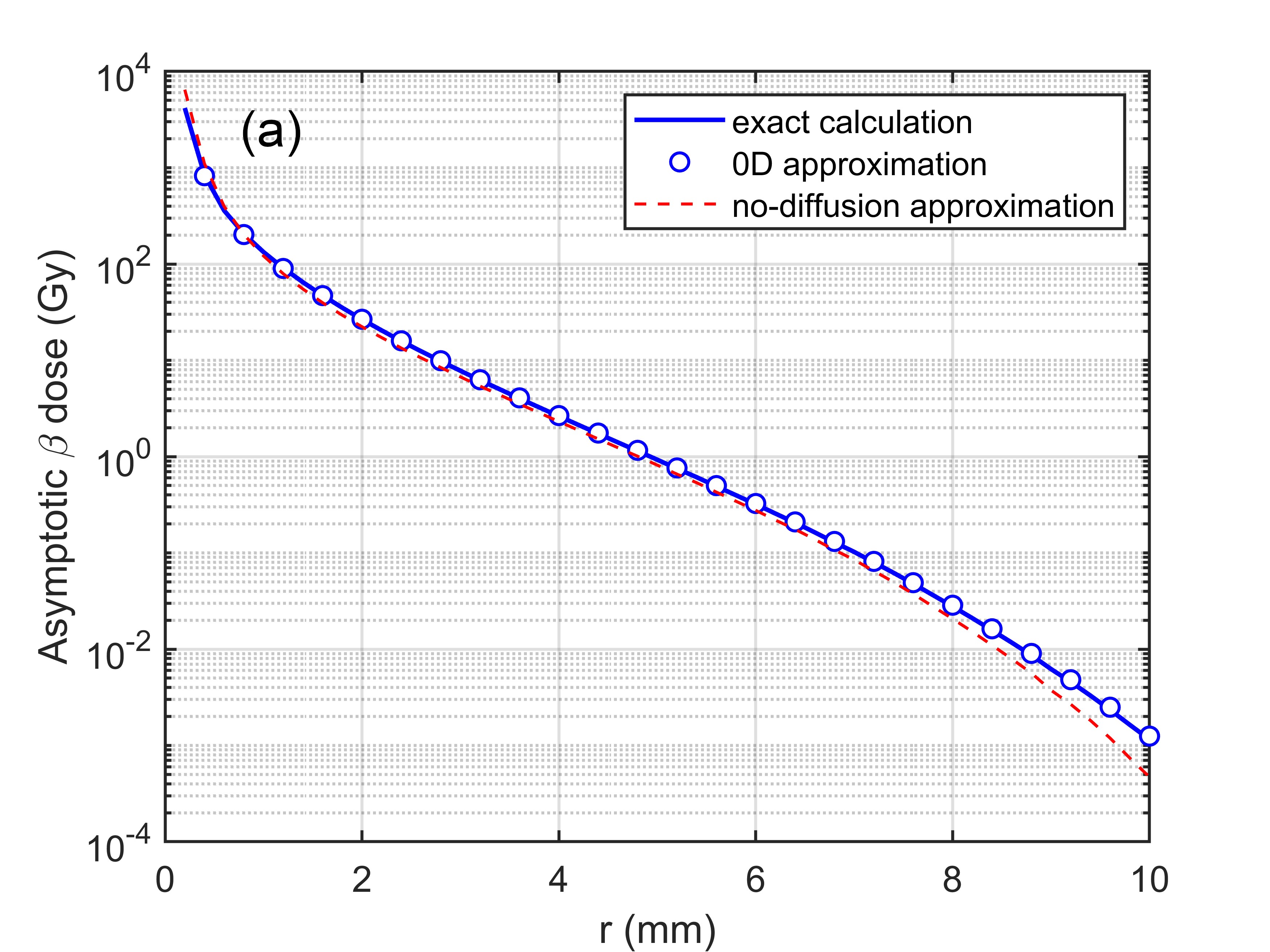}
	\includegraphics[scale=0.2,width=0.49\textwidth]{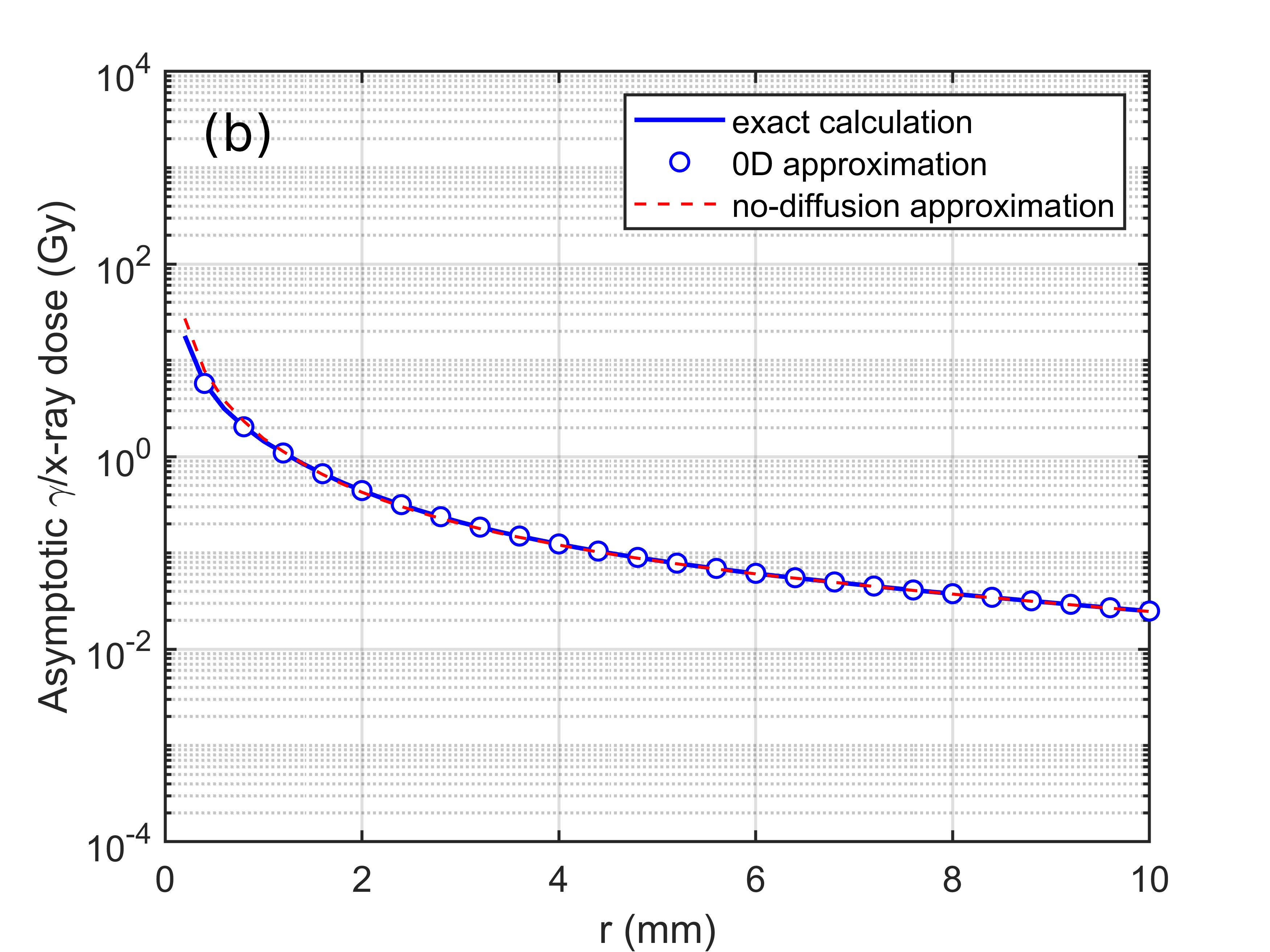}
	\caption{Asymptotic beta dose (a) and gamma/x-ray dose (b) for an Alpha DaRT point source, showing the exact solution, 0D approximation and ``no-diffusion'' approximation for a high-diffusion/low-leakage case. DPKs were calculated by EGSnrc. The source parameters are: $\Gamma_{Ra}^{src}(0) = 3~\mu$Ci, $P_{des}(Rn) = 0.45$, $P_{des}^{eff}(Pb) = 0.55$. The tumor tissue parameters are: $L_{Rn} = 0.3$~mm, $L_{Pb} = 0.6$~mm, $P_{leak}(Pb) = 0.3$.}
	\label{fig:Beta_and_gamma_dose_point_source}
\end{figure}

The advantage of the no-diffusion approximation is that it removes the need for modeling diffusion, and is therefore universal across all tumor types. The $^{212}$Pb leakage probability should still be considered and acts as an effective reduction in the source activity (limited to the contribution of the atoms outside the source). Figure~ \ref{fig:Beta_and_gamma_dose_point_source} compares the exact beta and gamma/x-ray asymptotic doses---calculated by numerically solving the diffusion-leakage equations and evaluating the integrals in Equations (\ref{eq:Dose_beta_point_source_triple_integral}) and (\ref{eq:Dose_gamma_point_source_triple_integral})---to the results of applying the 0D approximation\cite{Arazi2020} combined with the above integrals and to the results of the no-diffusion approximation. All calculations account for the source contribution, and were done using the EGSnrc DPKs. The dose was calculated for a point source with $\Gamma_{Ra}^{src}(0)=3~\mu$Ci and $P_{des}^{eff}(Pb)=0.55$. The diffusion lengths of $^{220}$Rn and $^{212}$Pb---which govern the spatial profile of the number densities and alpha dose\cite{Arazi2020}---were taken to represent a high-diffusion scenario with $L_{Rn}=0.3$~mm and $L_{Pb}=0.6$~mm, and the $^{212}$Pb leakage probability was set as $P_{leak}(Pb)=0.3$ to provide a low leakage level. Other parameters\cite{Arazi2020} were $P_{des}(Rn)=0.45$, $L_{Bi}=0.1\,L_{Pb}$ and $\alpha_{Bi}=0$. The 0D approximation of the beta and gamma/x-ray dose is accurate to $<1\%$ up to a few mm from the source ($\sim0.1\%$ at $r\sim2-3$~mm). The no-diffusion approximation underestimates the exact beta dose by $~14-17\%$ over the range $r=2-6$~mm and the exact gamma/x-ray dose by $\sim2-5\%$ over the same range. For a low-diffusion/high-leakage scenario, with $L_{Rn}=0.3$~mm, $L_{Pb}=0.3$~mm and $P_{leak}(Pb)=0.8$, underestimation of the beta dose drops to $\sim3\%$ for $r=2-6$~mm, while the gamma/x-ray dose is accurate to $\sim2\%$.

\subsection{Low-LET dose from a realistic Alpha DaRT source}
To estimate the low-LET dose for a realistic cylindrical Alpha DaRT source, including the contribution of diffusing atoms, we solved the 2D equations of the diffusion-leakage model numerically using the DART2D code\citep{Heger2023a}, to provide the total number of decays of $^{212}$Pb and $^{212}$Bi per unit volume throughout a cylindrical domain surrounding the source. These were then used to define the source location of beta and gamma/x-ray emissions in FLUKA. The 2D model was employed for two opposite cases mentioned above: 
\begin{enumerate}
    \item High-diffusion/low-leakage (``High spread'' scenario): $L_{Rn}$=0.3 mm,  $L_{Pb}$=0.6 mm, $P_{leak}(Pb)$=0.3.
    \item Low-diffusion/high-leakage (``Low spread'' scenario): $L_{Rn}$=0.3 mm,  $L_{Pb}$=0.3 mm, $P_{leak}(Pb)$=0.8.
\end{enumerate}
Other parameters used in both cases were: $L_{Bi}=0.1\,L_{Pb}$, $P_{des}(Rn)=0.45$ and $P_{des}^{eff}(Pb)=0.55$.

The MC simulation included the source geometry and material to account for interactions of the beta electrons and photons with the source structure. The source was taken as a stainless steel cylinder, with an outer radius of 0.35~mm and 10~mm in length. The simulation also comprised an inner poly-propane cylinder with a radius of 0.2~mm, representing a central suture used clinically to deploy chains of stranded sources. $^{224}$Ra and its daughter atoms remaining on the source were assigned uniformly to the source wall, with no activity on the cylinder bases, as is the case for real Alpha DaRT sources. Note that according to the manufacturer, variations in the activity density across the wall are below $\sim10\%$.

\begin{figure}
	\centering
	\includegraphics[width=0.49\textwidth]{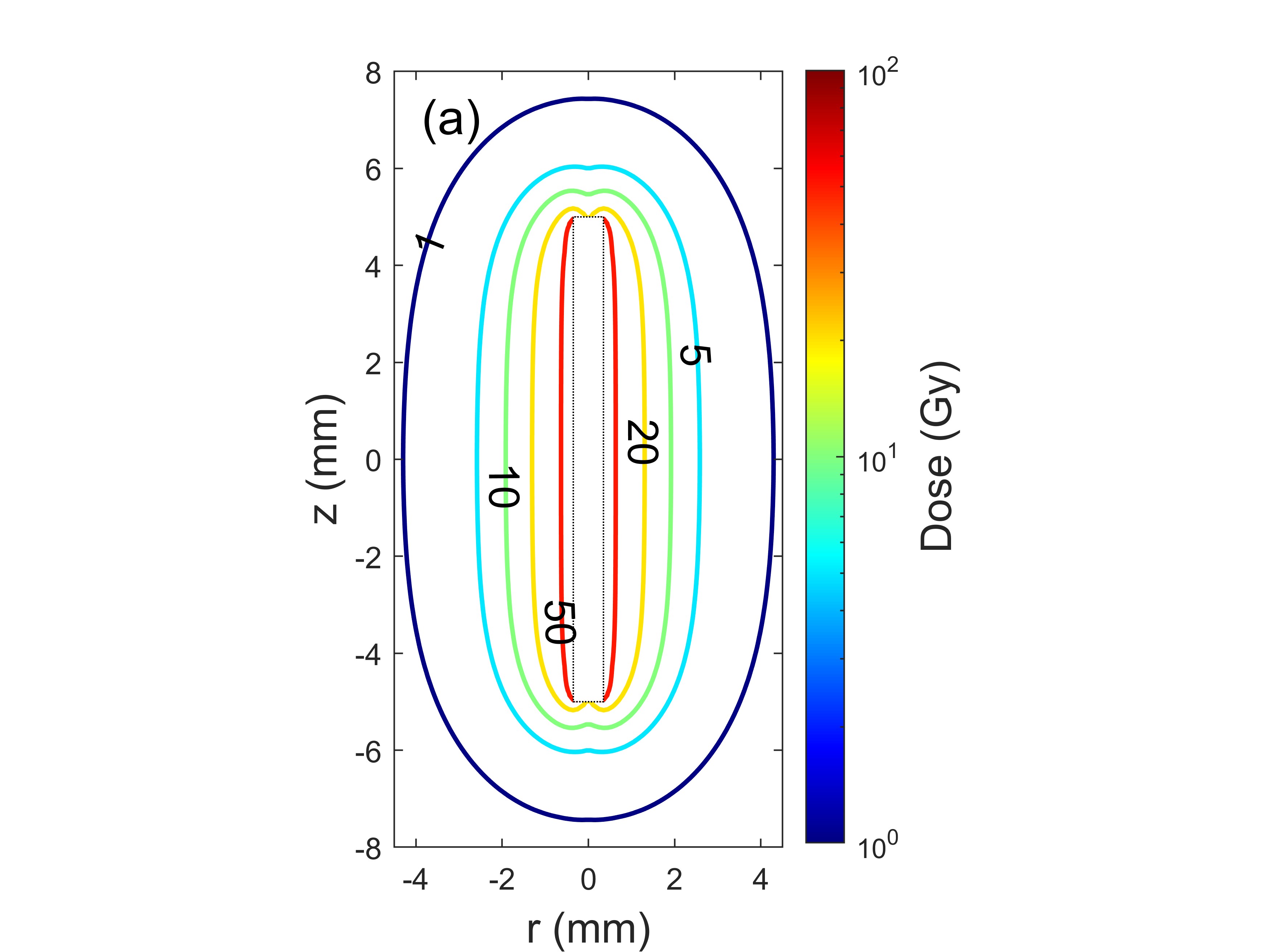}
	\includegraphics[width=0.49\textwidth]{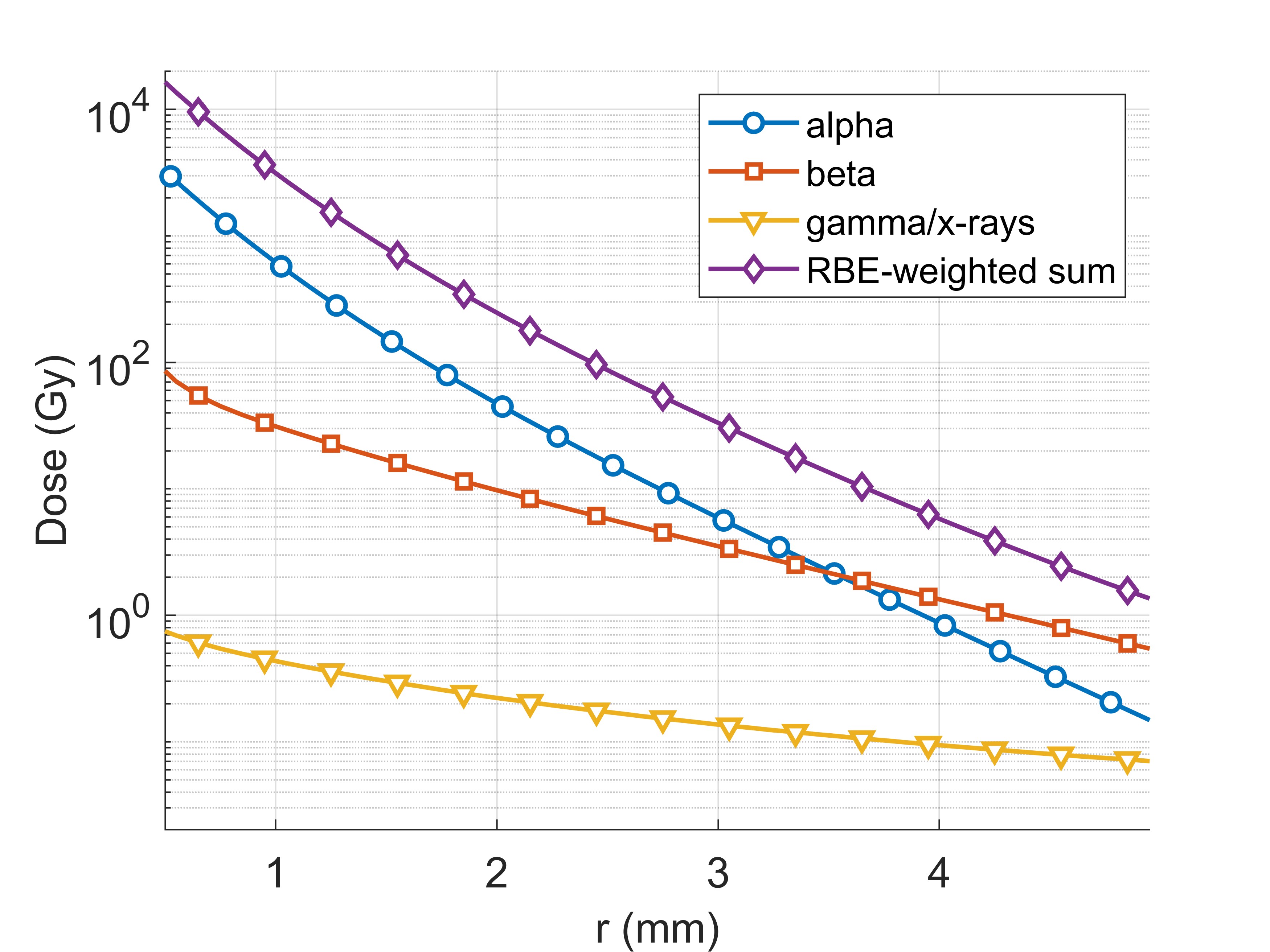}
	\caption{Dose distribution around a realistic source (a) Total asymptotic low-LET dose under high-diffusion/low-leakage conditions, as calculated using FLUKA for a spatial distribution of the diffusing atoms calculated using DART2D. (b) Asymptotic alpha, beta, gamma/x-ray, and RBE-weighted total dose as a function of distance in the source mid-plane; The source parameters are: $\Gamma_{Ra}^{src}(0) = 3~\mu$Ci, $P_{des}(Rn) = 0.45$, $P_{des}^{eff}(Pb) = 0.55$. The tumor tissue parameters are: $L_{Rn} = 0.3$~mm, $L_{Pb} = 0.6$~mm, $P_{leak}(Pb) = 0.3$.}
	\label{fig:2D dose plot}
\end{figure}

Figure \ref{fig:2D dose plot}a shows the results of the FLUKA calculation for the total asymptotic low-LET dose (electrons + photons) in the $rz$ plane of a realistic source with an initial $^{224}$Ra activity of 3~$\mu$Ci. Figure \ref{fig:2D dose plot}b shows the radial profiles of the asymptotic alpha, beta, and gamma/x-ray dose in the source mid-plane (the alpha dose was extracted directly from DART2D). The calculation was done for the high-spread scenario. The low-LET dose on the source surface is $>$100~Gy and is dominated by the beta contribution, as shown in figure \ref{fig:2D dose plot}b. Alpha particles are the main contributor to the dose below $\sim3$~mm, while the beta dose dominates at larger distances. Note that the alpha, beta, and gamma/x-ray curves represent the physical dose, without accounting for relative biological effectiveness (RBE). The RBE-weighted sum curve represents the total equivalent dose, accounting for an alpha-particle $RBE=5$ (as recommended by the MIRD 22 pamphlet \cite{MIRD_Pamphlet_22} for deterministic alpha-particle biological effects). The equivalent alpha dose is the dominant contributor up to $r\approx5$~mm.

\begin{figure}
    \centering
    \includegraphics[width=7.5cm]{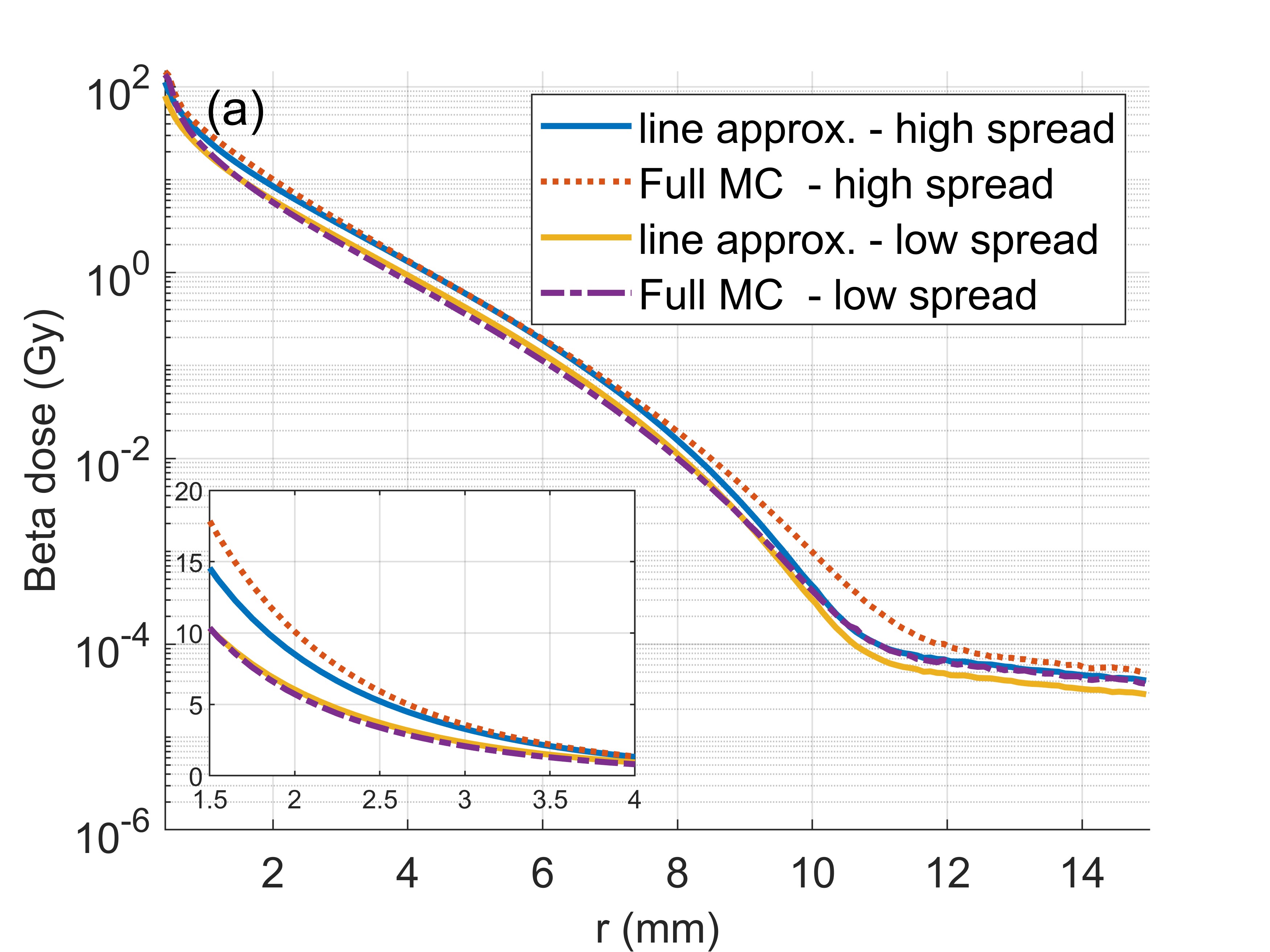}
    \includegraphics[width=7.5cm]{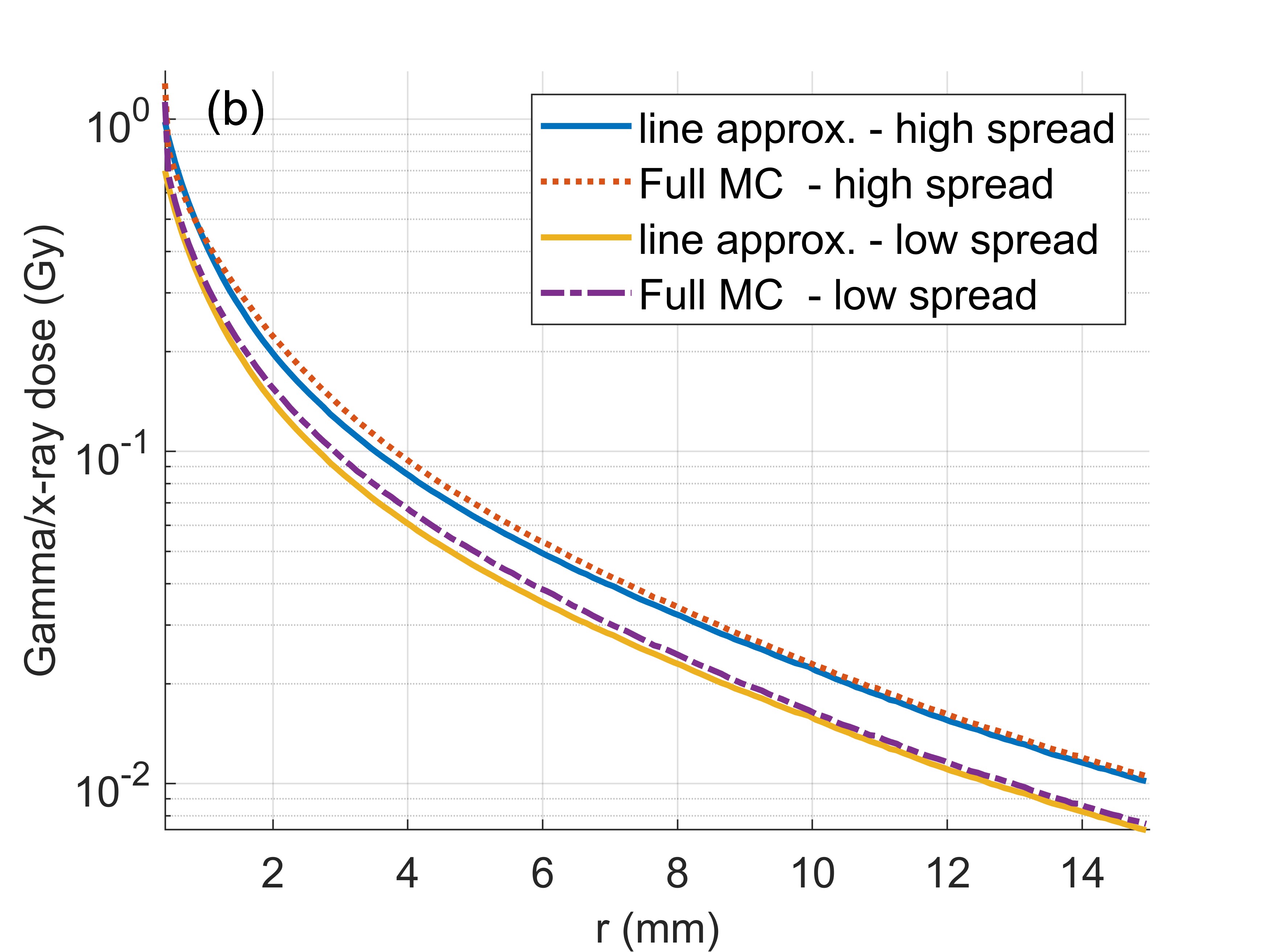}
    \caption{Total low-LET dose for the high-spread ($L_{Rn}=0.3$~mm, $L_{Pb}=0.6$~mm, $P_{leak}(Pb)=0.3$) and low-spread ($L_{Rn}=0.3$~mm, $L_{Pb}=0.3$~mm, $P_{leak}(Pb)=0.8$) scenarios. Comparison between the line-source/no-diffusion approximation and full MC calculation that includes a realistic source and the diffusing atoms. (a) Beta dose (with the inset focusing on the range 1.4-4~mm); (b) Gamma/x-ray dose. }
    \label{fig:DifDose}
\end{figure}

Figure \ref{fig:DifDose} shows a comparison between the full MC calculation for a realistic source with the diffusing atoms, and a line-source/no-diffusion approximation, calculated similarly to the point source in section \ref{sec:point source diffusing atoms}, where the line source is divided into point-like segments. The comparison is made for both the asymptotic beta dose (a) and gamma/x-ray dose (b). The doses were calculated for both the high- and low-spread scenarios. The comparison indicates a reasonable agreement between the line-source/no-diffusion approximation (which takes into account the correct $^{212}$Pb leakage probability) and the full MC calculation for $r\sim2-3$~mm, with better accuracy for the low-diffusion/high-leakage case.

The relative difference (underestimation) for the gamma/x-ray dose calculation is $\lesssim12\%$ for the high-diffusion/low-leakage case, and $\lesssim10\%$ for the low-diffusion/high-leakage scenario for $r>2$~mm, and decreases with increasing distance from the source. The line-source/no-diffusion approximation overestimates the beta dose by $\lesssim10\%$ for the low-spread case, and underestimates it by $\sim10-15\%$ for the high-spread scenario at $r=2-3$~mm.
The asymptotic beta dose at a distance of 2~mm from the source is 10.6~Gy under high-diffusion/low-leakage conditions and 6.0~Gy for the low-diffusion/high-leakage case, for an initial $^{224}$Ra activity of 3~$\mu$Ci. 

\section{Lattice calculations}
In this section we discuss the low-LET dose in a hexagonal DaRT source lattices, which---as shown previously\cite{Heger2023b}---provide the optimal alpha dose coverage for a given source spacing inside the tumor. As for the alpha dose, the dose at any point inside and outside of the lattice is a linear superposition of contributions from all sources. However, unlike the alpha dose, whose rapid radial fall-off results in negligible contribution from sources beyond the three nearest ones, the beta and gamma dose profiles drop more slowly, and additional sources contribute to the dose at any point.

\begin{figure}
    \centering
    \includegraphics[width=7.5cm]{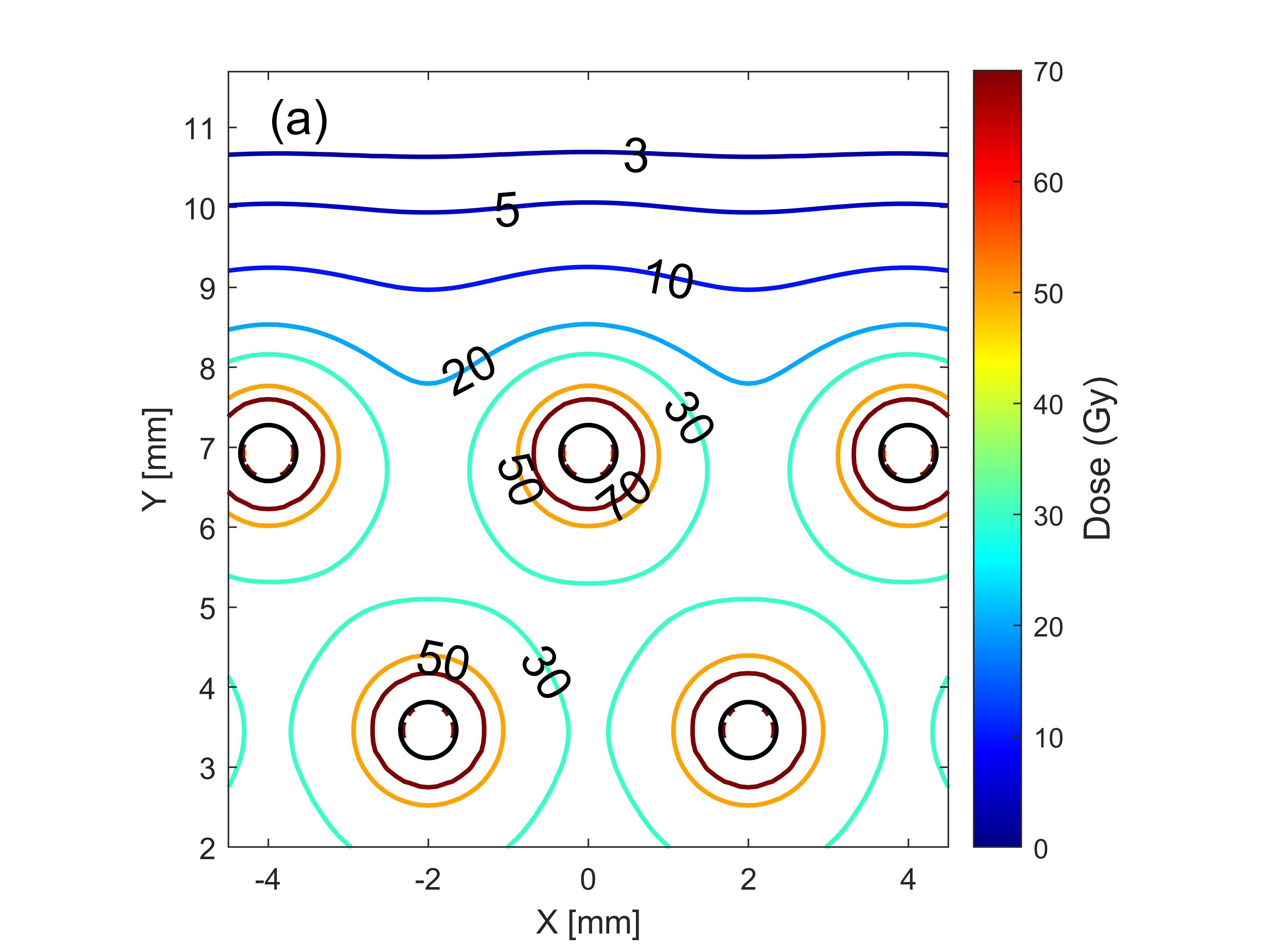}
    \includegraphics[width=7.5cm]{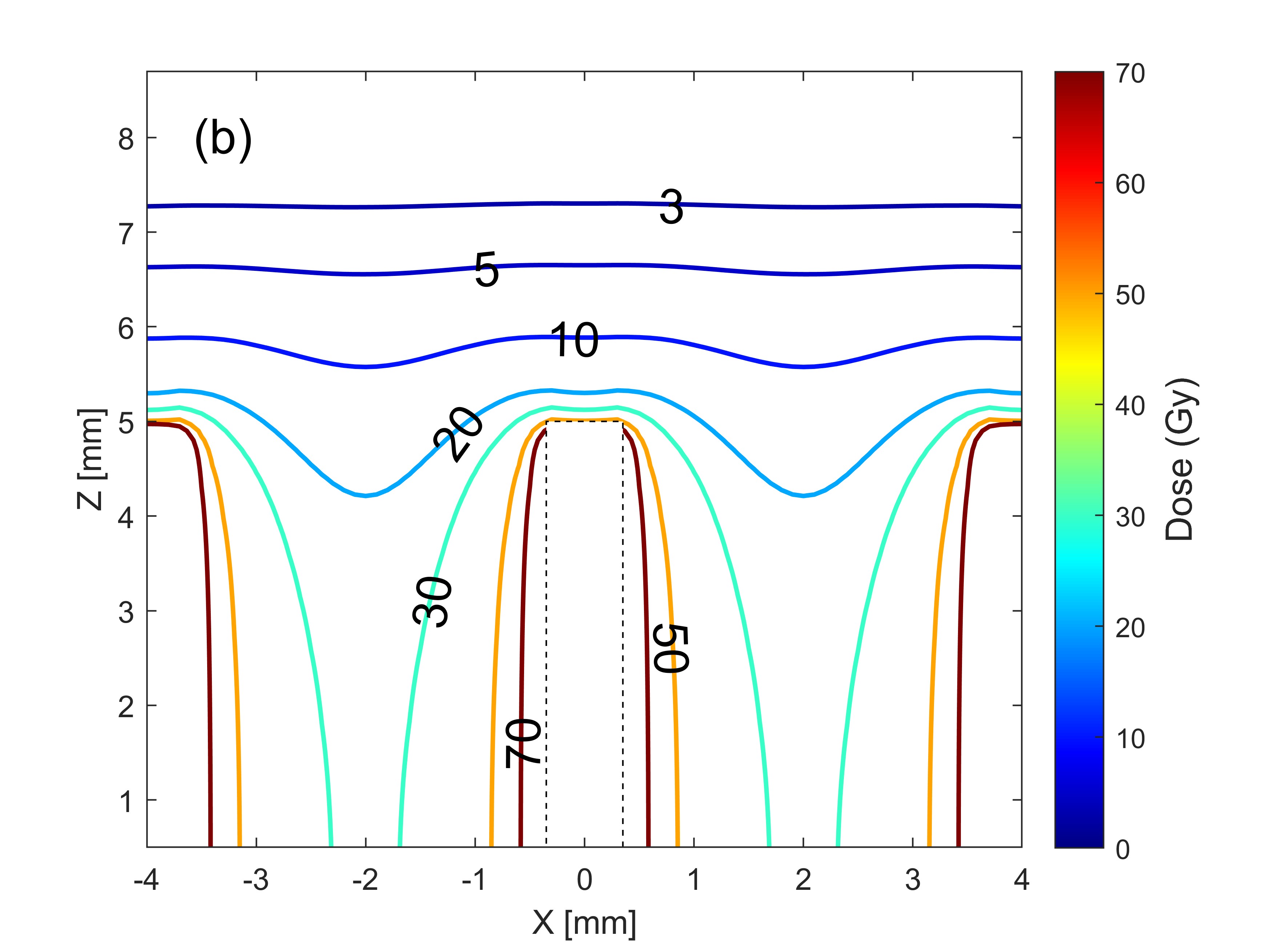}
    \caption{Total dose distribution for high diffusion ($L_{Rn} = 0.3$~mm, $L_{Pb} = 0.6$~mm) and low leakage ($P_{leak}(Pb)=0.3$) conditions in lattice formation. (a) axial view at the sources mid-plane; (b) side view. }
    \label{fig:laticeDose}
\end{figure}

Figure \ref{fig:laticeDose} shows the asymptotic low-LET dose in a $5\times5$ hexagonal lattice comprising sources carrying $3~\mu$Ci/cm $^{224}$Ra with 4~mm spacing for the high-diffusion/low-leakage case. The dose calculations include the contribution of all sources in the lattice. Figure \ref{fig:laticeDose}(a) shows the dose distribution across the source midplane and (b) displays it in a plane containing the source axes. The dose to surrounding tissues outside of the lattice drops to $<5$~Gy at a distance of 3 mm away from the outermost source. We define the effective initial dose rate as the asymptotic dose divided by the mean lifetime of $^{224}$Ra. We use the term ``effective'' since the low-LET dose rate from the diffusing atoms builds up from zero to a maximal value (around 1 day after source insertion), before reaching secular equilibrium with the source. With this definition, the effective initial dose rate 3~mm away from the outermost source is $<0.04$~Gy/h.

\begin{figure}
    \centering
    \includegraphics[width=0.35\paperwidth]{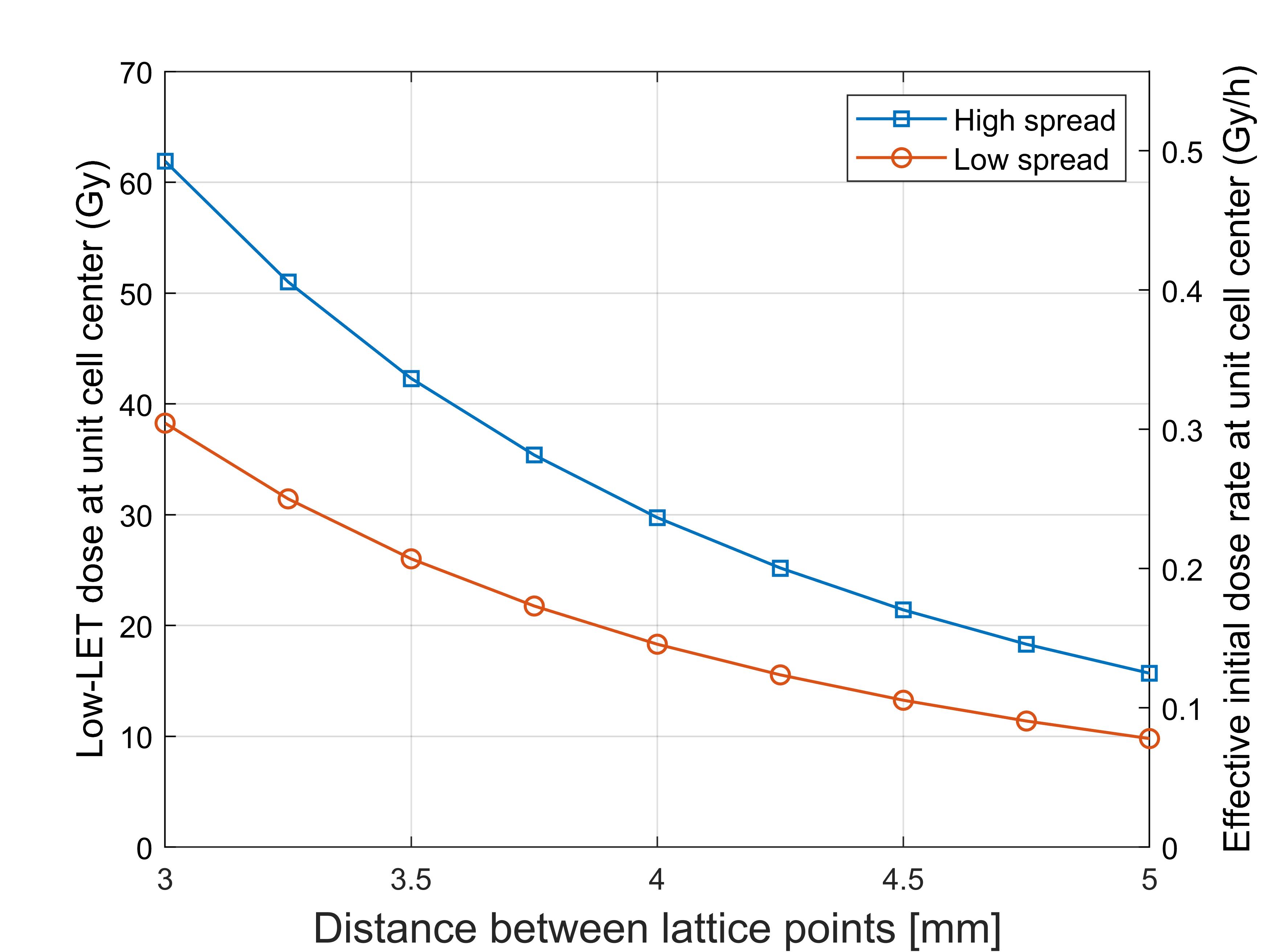}
    \caption{Asymptotic low-LET dose (left axis) and effective initial dose rate (right axis) at the center between three adjacent sources deep inside a large hexagonal lattice for the high-diffusion/low-leakage and low-diffusion/high-leakage scenarios.}
    \label{fig:diflaticedistDose}
\end{figure}

Figure \ref{fig:diflaticedistDose} shows the low-LET asymptotic dose (left axis) and effective initial dose rate (right axis) at the center-of-gravity between three adjacent sources deep inside a hexagonal lattice as a function of the source spacing for the low- and high-spread conditions defined above. At 4~mm spacing, the minimal low-LET dose is $\sim$30~Gy for the high-spread case and 18~Gy for low-spread conditions. For random source placement errors on the scale of $\sim0.5$~mm, the relative change in the minimal beta dose between three sources is $\sim20-30\%$. The effective initial dose rate is on the same scale as that of conventional low-dose-rate brachytherapy treatments (for example, 0.21~Gy/h with $^{103}$Pd and 0.07~Gy/h with $^{125}$I treatments for prostate cancer\cite{Tang2023}.)

\section{Effect of the low-LET dose on cell survival}
Cell survival and tumor control probability (TCP) modeling in Alpha-DaRT are nontrivial, because of the strong spatial variation of the tumor dose and the stochastic nature of alpha-particle hits to cell nuclei. A thorough study of Alpha-DaRT microdosimetry is presently ongoing and deferred to a separate publication. However, one can estimate the contribution of the low-LET dose to cell survival with three simplifying assumptions: (1) it can be modeled macroscopically without considering stochastic effects; (2) it acts independently of the alpha dose; (3) it is most important to consider cell survival in the tumor regions subject to the lowest radiation doses.

Assumption (1) can be justified by considering the typical number of beta-electron trajectories crossing the cell nucleus. Consider a spherical nucleus with a radius $R_{nuc}=2.5~\mu$m. The average chord length crossing the nucleus is $\bar l=4R/3=3.3~\mu$m \cite{Brantley2011} and the nucleus mass (assuming liquid water density) is $M_{nuc}=6.5\cdot10^{-14}$~kg. Assuming $dE/dX=0.2$~keV/$\mu$m, the average dose contributed by one beta electron crossing the nucleus is $\bar d=dE/dX\cdot\bar l/M_{nuc}=1.6\cdot10^{-3}$~Gy. If the total low-LET absorbed dose in this region is 25~Gy, the mean number of electron hits to the nucleus is therefore $\sim1.5\cdot10^4$, with a relative standard deviation of 0.8\%. Thus, stochastic variations in the low-LET dose are on the percent level and can be expected to have a second-order effect on cell survival.

Assumption (2) means that the low-LET dose acts on cells that are either already dead (i.e., have lost their multiplicative potential) or free of any DNA damage. Mathematically, it can be expressed as: $SF(D_{tot};independent)=SF(D_{\alpha})\cdot SF(D_{\beta+\gamma})$ (where $SF$ is the surviving fraction). In a true mixed-field analysis, one should obviously consider the case where DNA damage induced by low-LET electrons crossing a given nucleus adds up to existing sublethal damage by alpha particles (and vice versa) \cite{GuerraLiberal2023, Cheng2018}. This would result in lower survival: $SF(D_{tot};mixed~field)<SF(D_{\alpha})\cdot SF(D_{\beta+\gamma})$. Thus, the assumption of independent action of the high-LET and low-LET components underestimates the true cell-killing effect of the mixed field and is therefore conservative. 

Assumption (3) stems from the rapid spatial variation of the alpha dose with the distance from the source within source lattices. For alpha particles, the regions of minimal dose between three adjacent sources dominate the overall cell survival and TCP. Even a sub-mm shift from these points increases the local alpha dose by a large factor, dramatically reducing local cell survival. Since the low-LET dose attains its minimal value (inside the lattice) at the same locations as the alpha dose, these are the points where its contribution to the TCP is most important.

Under these assumptions, and since Alpha-DaRT is a form of protracted radiotherapy, the local surviving fraction arising from the action of the low-LET dose is given by the  linear-quadratic (LQ) model with a dose-rate correction factor $q(t)$:
\begin{equation}
    SF(D_{\beta+\gamma}(t))=e^{-\alpha D_{\beta+\gamma}(t)-\beta q(t) D_{\beta+\gamma}^2 (t)}   \label{eq:LQ quadratic}
\end{equation}
with $t$ representing the time from source insertion. For a protracted therapy, where the dose rate decays proportionally to $e^{-\lambda t}$, $q(t)$ is given by \cite{Wuu_Zaider1994}:
\begin{equation}
    q(t)=\frac{2(\lambda/\mu)^2}{ \left(1-(\lambda/\mu)^2 \right) (1-e^{-\lambda t}) }\left[ e^{-(\lambda+\mu)t} + \frac{\mu}{2\lambda}\left( 1-e^{-2\lambda t} \right) - \frac{1+e^{-2\lambda t}}{2} \right] \label{eq:q(t)}
\end{equation}
where $\mu$ is the DNA damage repair rate (inverse of the mean time for repair, $\sim0.5-5$~h, i.e. $\mu\sim0.2-2$~h$^{-1}$)\cite{Bentzen1999, Hall2018, Shaikh2020}.

In Alpha-DaRT, if the source is in secular equilibrium at insertion, its contribution to the low-LET dose rate is proportional everywhere to $e^{-\lambda_{Ra}t}$. The low-LET dose rate arising from the diffusing atoms reflects the initial buildup phase of $^{212}$Pb. Under the 0D time-dependence approximation\cite{Arazi2020} it is proportional everywhere to $\left(e^{-\lambda_{Ra}t}-e^{-(\lambda_{Pb}+\alpha_{Pb})t}\right)$, where $\alpha_{Pb}$ is the clearance rate coefficient of $^{212}$Pb by the blood, related to the leakage probability by $P_{leak}(Pb)=\alpha_{Pb}/(\lambda_{Pb}+\alpha_{Pb})$. For $P_{leak}(Pb)=0.5$, $\lambda_{Ra}/(\lambda_{Pb}+\alpha_{Pb})=0.06$, so that up to $\lesssim10\%$ error, the dose rate from the diffusing atoms is also proportional to $e^{-\lambda_{Ra}t}$. Therefore, one can adopt Equation (\ref{eq:q(t)}) for $q(t)$ in Alpha-DaRT, with $\lambda=\lambda_{Ra}$.

\noindent For a treatment duration of several weeks, $\lambda_{Ra}t\gg1$. In this limit:
\begin{equation}
    \lim_{\lambda t \to \infty} q(t) = q(\infty) = \frac{\lambda}{\lambda+\mu}    
\end{equation}

Taking, for example, $\mu=1~$h$^{-1}$, this gives $q(\infty)=0.008$. For a typical value of $\alpha/\beta=10$~Gy and a low-LET asymptotic dose $D_{\beta+\gamma}=25$~Gy, the ratio between the quadratic and linear terms in Equation (\ref{eq:LQ quadratic}) is, in this case, $\beta q(\infty) D_{\beta+\gamma}^2/\alpha D_{\beta+\gamma}=0.02$. Therefore, the quadratic term has a second-order effect on the asymptotic surviving fraction of the low-LET dose, and:
\begin{equation}
    SF(D_{\beta+\gamma}^{asy})\approx e^{-\alpha D_{\beta+\gamma}^{asy}}   \label{eq:LQ linear}    
\end{equation}

Compilations of clinical $\alpha$ values\cite{VanLeeuwen2018} indicate a typical range of $\sim0.02-0.2$~Gy$^{-1}$ with some dependence on tumor type. For a low-LET asymptotic dose of 25~Gy, this gives $SF(D_{\beta+\gamma}^{asy})\sim0.007-0.6$. The tumor control probability (i.e., the probability that no clonogenic cells survive the treatment) is given by\cite{Munro1961,Roeske2000}:
\begin{equation}
    TCP=e^{-\bar N_{survive}}
\end{equation}
where $\bar N_{survive}$ is the average number of surviving clonogenic cells expected in the treatment. Consider a case where, in the absence of the low-LET dose, $\bar N_{survive}=0.69$ and $TCP=0.5$. For a low-LET surviving fraction in the range $0.007-0.6$, the inclusion of the low-LET dose gives $\bar N_{survive}=0.005-0.4$, increasing the TCP to $0.76-0.997$. This shifts the TCP curve by up to a few Gy to lower values relative to the alpha-only case, where the dose corresponding to $TCP=0.5$ is $\sim10-20$~Gy, depending on the nucleus size, cellular radiosensitivity and tumor volume\cite{Roeske2000}. Preliminary calculations, to be discussed in detail in a separate publication, indicate that the reduction in the required alpha dose is more pronounced for small nuclei with high radiosensitivity. 

\section{Summary and discussion}

Alpha DaRT has shown promising results in treating solid tumors. This success is attributed primarily to the migration of the alpha-emitting atoms, which creates a high-LET, high-dose region over a few mm around each source, and was the subject of several previous publications. Here we provide a first in-depth discussion of the non-negligible low-LET dose accompanying the Alpha DaRT treatment.

We first examined the isotope-specific beta and x-ray/gamma dose point kernels (DPKs) of $^{224}$Ra and its daughters, which we calculated using the EGSnrc and FLUKA MC codes. We showed that the most significant contributions to the low-LET dose at therapeutically relevant distances from the source ($r\sim2-3$~mm) come from the beta decays of $^{212}$Bi and $^{208}$Tl. The x-ray/gamma dose is governed by the 2.615~MeV gamma of $^{208}$Tl and, over this range, is $>30$-fold smaller than the beta dose. We found good agreement between the two codes, particularly for the leading dose contributions. Differences between the codes can be attributed, at least in part, to energy cutoffs. The cutoffs for electrons and photons were selected as 10~keV and 1~keV respectively for both codes. While this applies to particle production in the two codes, the EGSnrc cross-section libraries that were used have a lower limit of 10~keV for both electrons and photons. Thus, particles below 10~keV are forced to deposit their energy locally in EGSnrc. This should result in higher energy depositions closer to the source compared to the FLUKA results. Larger differences between the codes were indeed observed in the dose from decay products with lower energies and especially energies that are closer to the cutoff energies, for example, the 11~keV photons emitted from $^{212}$Pb with 14.3\% emission probability.

Next, we considered the total beta dose and x-ray/gamma dose from an Alpha DaRT point source, accounting separately for electron and photon emission from atoms located on the source itself, and from atoms diffusing in the source vicinity. The ratio between the contribution of the diffusing atoms and that of the source is roughly $1-P_{leak}(Pb)$, and thus for typical values of the $^{212}$Pb leakage probability ($\sim0.3-0.7$), the source contributes $\sim60-80\%$ of the total the low-LET dose. The source contribution depends on the ratio between the $^{212}$Pb and $^{224}$Ra activities it carries at the time of treatment. Here, variations in the amount of $^{212}$Pb activity contained in the glycerin surrounding a real source can result in $\sim5\%$ variations in the total low-LET dose. Concerning the low-LET dose contributed by the diffusing atoms, we showed that ``switching off'' their diffusion and assuming that all their beta and x-ray/gamma emissions originate from the source itself provides a reasonable first-order approximation when $^{212}$Pb leakage is properly accounted for. This holds because the range of the beta electrons emitted by $^{212}$Bi and $^{208}$Tl (and clearly the mean free path of the 2.615~MeV gamma ray of $^{208}$Tl) is much larger than the diffusion lengths governing the spread of the diffusing atoms.

After studying the different factors affecting the low-LET dose, we moved on to describe a realistic geometry of an Alpha DaRT source. We employed a full MC calculation in FLUKA, including contributions from atoms residing on the source surface and from atoms diffusing around the source. The latter was found by solving numerically the time-dependent diffusion-leakage model equations to provide the starting point for the emission of beta electrons and x-ray/gamma photons. The calculation was done for both high-diffusion/low-leakage and low-diffusion/high-leakage cases. We showed that a line-source/no-diffusion approximation provides an accuracy of $\sim10-15\%$ for the total low-LET dose at therapeutically-relevant distances for both cases.

For Alpha DaRT source lattices, we found that the minimal low-LET dose between three adjacent $3~\mu$Ci/cm $^{224}$Ra sources in a hexagonal lattice with 4~mm spacing is $\sim30$~Gy for the high-diffusion/low-leakage scenario and $\sim18$~Gy for the low-diffusion/high-leakage case. The minimal dose changes by $\sim20-30\%$ for $\pm0.5$~mm changes in the source spacing. The low-LET dose drops below 5~Gy $\sim3$~mm away from the outermost source in the lattice, with an effective maximal dose rate of $<0.04$~Gy/h, ensuring the sparing of healthy tissue (at the same distance the alpha dose is typically below $\sim1$~Gy).

A minimal nominal dose of 25~Gy in a 4~mm lattice with 3~$\mu$Ci/cm sources can help reduce cell survival by a significant factor. Since the dose rate is very low, one can neglect the quadratic term in the linear-quadratic model, and the low-LET-associated survival reduction factor is $SF(D_{\beta+\gamma})=e^{-\alpha D_{\beta+\gamma}}$. With clinical $\alpha$ values in the range $\sim0.02-0.2$~Gy$^{-1}$, $S(D_{\beta+\gamma})\sim0.007-0.6$. This, in turn, can shift tumor control probability curves to lower alpha doses by up to a few Gy.

Further significant improvement can be obtained by increasing the $^{224}$Ra activity on the source by, for example, a factor of 5 (to $15~\mu$Ci) while reducing the desorption probabilities of $^{220}$Rn and $^{212}$Pb by the same amount. This will drive the minimum low-LET dose between sources to therapeutic levels. It will further allow increasing the lattice spacing while maintaining the total equivalent dose inside the tumor and the same level of dose to distant organs (due to leakage of $^{212}$Pb into the circulation\cite{Arazi2010}). As an example, consider a low-diffusion/high-leakage case ($L_{Rn}=0.3$~mm, $L_{Pb}=0.3$~mm, $P_{leak}(Pb)=0.8$), where the spacing required for tumor control in a particular tumor type with standard $3~\mu$Ci/cm $^{224}$Ra Alpha DaRT sources is known to be 4~mm. Using DART2D, the minimal alpha dose between three sources, in this case, is 13.9~Gy, and from Figure \ref{fig:diflaticedistDose} the beta dose at the same point is 18~Gy. If we assume that for alpha particles $RBE=5$, the equivalent total dose is 88~GyE. From Equations (\ref{eq:BetaDoseSrc_point_source}) and (\ref{eq:Dose_beta_no_diff}) the total low-LET dose is roughly proportional to $\Gamma_{Ra}^{src}(0)\cdot\left(1-P_{des}^{eff}(Pb)P_{leak}(Pb)\right)$. Setting $\Gamma_{Ra}^{src}(0)=15~\mu$Ci, $P_{des}(Rn)=0.09$ and $P_{des}^{eff}(Pb)=0.11$ will therefore boost the low-LET dose by a factor of $\sim8$.  With the boosted low-LET dose the spacing can now be increased to 5~mm to maintain the same equivalent tumor dose: the minimal alpha dose between three sources at 5~mm spacing is 2.0~Gy and the minimal boosted beta dose is $\sim9.8$~Gy$\times8=78.4$~Gy, so for $RBE=5$ the total equivalent dose is again $\sim88$~GyE. Note that a boosted low-LET dose can also be used to improve tumor control in case it is not sufficiently high with standard Alpha DaRT sources, by keeping the same lattice spacing, or increasing it by a smaller amount than suggested above. Additionally, a boosted low-LET dose can potentially improve tumor control in cases where hexagonal source placement is not practical. On the other hand, the low-LET dose in a boosted scenario will extend further (by $\sim2-3$ mm) into the surrounding healthy tissue---e.g., dropping to 5~Gy $\sim5$~mm away from the outermost source in a 5~mm lattice for a 5-fold boost in activity. Clearly, such calculations are only indicative in nature (in particular, with respect to the actual RBE), and should only be considered as an aid to choosing the starting point for clinical trials in terms of source activity and nominal spacing.

\section{Conclusion}
In this work, we provided the first discussion of the low-LET dose in Alpha DaRT. We showed that the low-LET dose is dominated by the beta emissions of $^{212}$Bi and $^{208}$Tl (from both the source and diffusing atoms, where the source contribution is typically higher), with a negligible effect of x- and gamma rays. Depending on the diffusion and leakage conditions, the minimal low-LET dose inside a hexagonal lattice of $3~\mu$Ci/cm $^{224}$Ra sources with 4~mm spacing is $\sim18-30$~Gy, which helps to reduce cell survival by a considerable factor. The low-LET dose drops to negligible levels already $\sim3$~mm away from the treated region, ensuring the sparing of surrounding healthy tissue. Increasing the source activity by, for example, a factor of 5, while reducing the desorption probabilities of $^{224}$Ra daughters from the source by the same amount, can boost the low-LET dose itself to therapeutic levels, potentially allowing a high probability of tumor control with an increased source spacing.

\section*{References}
\addcontentsline{toc}{section}{\numberline{}References}
\vspace*{-20mm}

\bibliography{references}
\bibliographystyle{./medphys.bst}

\clearpage

\end{document}